\documentclass[a4paper,twocolumn,pra,superscriptaddress]{revtex4}
\usepackage[english]{babel}

\usepackage{amssymb,amsthm,mathrsfs,amscd,amsmath}
\usepackage{multirow}
\usepackage{array}

\usepackage{graphicx,color}

\DeclareMathOperator{\e}{e} 
\newcommand{\ket}[1]{{| #1\rangle}}
\newcommand{\bra}[1]{{\langle #1|}}
\makeatletter
\newsavebox{\@brx}
\newcommand{\llangle}[1][]{\savebox{\@brx}{\(\m@th{#1\langle}\)}%
  \mathopen{\copy\@brx\kern-0.5\wd\@brx\usebox{\@brx}}}
\newcommand{\rrangle}[1][]{\savebox{\@brx}{\(\m@th{#1\rangle}\)}%
  \mathclose{\copy\@brx\kern-0.5\wd\@brx\usebox{\@brx}}}
\makeatother

\begin{document}
\title{Topological two-body bound states in the interacting Haldane model}

\author{Grazia Salerno}
\affiliation{INO-CNR BEC Center and Dipartimento di Fisica, Universit\`a di Trento, I-38123 Povo, Italy}
\affiliation{Center for Nonlinear Phenomena and Complex Systems, Universit\'{e} Libre de Bruxelles, CP 231, Campus Plaine, B-1050 Brussels, Belgium}
\author{Marco Di Liberto}
\affiliation{INO-CNR BEC Center and Dipartimento di Fisica, Universit\`a di Trento, I-38123 Povo, Italy}
\affiliation{Center for Nonlinear Phenomena and Complex Systems, Universit\'{e} Libre de Bruxelles, CP 231, Campus Plaine, B-1050 Brussels, Belgium}
\author{Chiara Menotti}
\affiliation{INO-CNR BEC Center and Dipartimento di Fisica, Universit\`a di Trento, I-38123 Povo, Italy}
\author{Iacopo Carusotto}
\affiliation{INO-CNR BEC Center and Dipartimento di Fisica, Universit\`a di Trento, I-38123 Povo, Italy}

\date{\today}

\begin{abstract}
We study the topological properties of the two-body bound states in an interacting Haldane model as a function of inter-particle interactions. In particular, we identify topological phases where the two-body edge states have either the same or the opposite chirality as compared to single-particle edge states. We highlight that in the moderately-interacting regime, which is relevant for the experimental realization with ultracold atoms, the topological transition is affected by the internal structure of the bound state and the phase boundaries are consequently deformed.
\end{abstract}

\maketitle

\section{Introduction}
The experimental advances and the deepened theoretical understanding in ultracold atoms \cite{Atala2013, Aidelsburger2013, Miyake2013, Jotzu2014, Mancini2015, Stuhl2015}, photonics \cite{Hafezi2013, Rechtsman2013, Chen2014, Ningyuan2015, Wang2009} and phononic systems \cite{Susstrunk2015, Nash2015, Wang2015, He2016} have made possible the implementation of topologically non-trivial models on the lattice. As these setups are made of neutral constituents, it has been necessary to develop schemes for the realization of artificial gauge potentials \cite{Dalibard2011, Aidelsburger2011, Goldman2014, Aidelsburger2017}. Single-particle physics has been widely explored in these platforms, thus opening the way to the investigation of the more demanding many-body regime. The interplay between interactions and topology indeed holds the promise to discover exotic phases of matter, as the famous fractional quantum Hall effect \cite{Tsui1982, Laughlin1983, Sorensen2005, Hafezi2007}, or the even more challenging symmetry-protected topological phases (SPTs) \cite{Chen2012}.

In preparation to the full many-body case, a simpler but nevertheless rich scenario in which to explore the role of interactions in topological systems is the few-body regime. 
Two interacting particles on a lattice can form a bound state (doublon) for both repulsive and attractive interactions, as a consequence of the finite single-particle bandwidth in discrete models \cite{Winkler2006,Valiente2008,Valiente2009}.
The doublon wavefunction is very localized in the relative coordinate, but the doublon center-of-mass can move across the lattice through higher-order tunneling processes. In this sense, doublons behave as particles with a large effective mass \cite{Winkler2006,Mukherjee2016}. 
However, their composite nature arises dramatically, \textit{e.g.} in affecting the boundary conditions in an open system, leading to interaction-induced Tamm-Shockley localization of the doublon at the edges \cite{Longhi2013,DiLiberto2016,DiLiberto2017,Gorlach2017,Gorlach2017b}. 

Most recently, increasing theoretical interest has been devoted to the understanding of the effect of topology in two-body systems \cite{Lim2011,DiLiberto2016,Gorlach2017,Bello2016,Bello2017,Marques2017,Marques2017b,Lee1,Lee2}. In one dimensional models such as the Su-Schrieffer-Heeger model \cite{DiLiberto2016,Gorlach2017,Bello2016}, on-site interactions break chiral symmetry that protects the topological edge states, so that the presence of two-body topological edge states is no longer ensured. Conversely, in two dimensions, topological properties can appear also when all the symmetries are broken (\emph{e.g.} Chern insulators) and interactions can therefore play an active role to induce novel and different topological transitions. 

Recent experiments observing chiral properties of two-body states have addressed the dynamics of two interacting photons in a single triangular plaquette \cite{Google} and two interacting bosonic atoms on a ladder \cite{Greiner2017} with a non-vanishing flux. 
Realizations of 2D geometries with a large number of sites in each dimension are going to be within experimental reach in the near future and theoretical studies in this direction are therefore of great interest.

A first theoretical attempt to study two-body topological systems in 2D is represented by the hard-core two-magnon excitations of a spin model on a square lattice with finite flux, anisotropic hopping and nearest-neighbour interactions \cite{Lee1,Lee2}. However, such a rather speculative model lacks a straightforward experimental implementation with current technologies. We instead consider the Haldane model \cite{Haldane1988} as a minimal and experimentally realistic Hamiltonian containing the main remarkable topological ingredients. This model has been already realized in the non-interacting regime with ultracold atoms \cite{Jotzu2014}. Moreover, the dynamics in the lattice is accessible through single-site manipulation \cite{Bakr2009, Greiner2017} and protocols to observe topological edge states have been proposed \cite{Goldman2013,Goldman2016}.

In this work, we investigate the topological properties of two interacting bosonic atoms in the Haldane model with on-site interactions. 
Our analysis shows that in the limit of very strong interactions, the doublon effectively behaves like a single-particle with renormalized parameters. 
We identify regimes where the doublon topological edge states exhibit either the same or the opposite chirality compared to the single-particle topological edge states.
Even more interestingly, for the experimentally relevant regime of moderate interactions, the composite nature of the doublons is responsible for a sizeable modification of the topological phase diagram. 
Whereas we focus our presentation on the case of two bosonic particles for simplicity, our results can be extended to a pair of distinguishable particles irrespectively of their statistics.

The paper is organized as follows. In Sec.~\ref{sec:model}, we introduce the interacting Haldane model. 
In Sec.~\ref{sec:effectiveHam}, we derive an effective Hamiltonian for the two-body bound states, which applies from the strongly- to the moderately-interacting regime.
In Sec.~\ref{sec:doublon}, we discuss the doublon spectrum, which is obtained from the effective Hamiltonian and from exact-diagonalization calculations. We identify topological edge states, as well as non-topological Tamm-Shockley edge states. 
In Sec.~\ref{sec:topo}, the topological properties of the doublons are characterized. We investigate the topological phase diagram from the strongly- to the moderately-interacting regime, and we discuss how the phase boundaries are modified by the interactions. In particular, we compare the chirality of the doublon edge states with the one of the single particle states in regimes of interactions that could be accessible in ultracold atom experiments.
In Sec.~\ref{sec:smallU}, we briefly discuss the regime of small interactions, where the doublon bands approach the scattering continuum.
Finally, we draw our conclusions in Sec.~\ref{sec:conclusions}.

\begin{figure}[t]
\centering
\includegraphics[width=0.48 \textwidth]{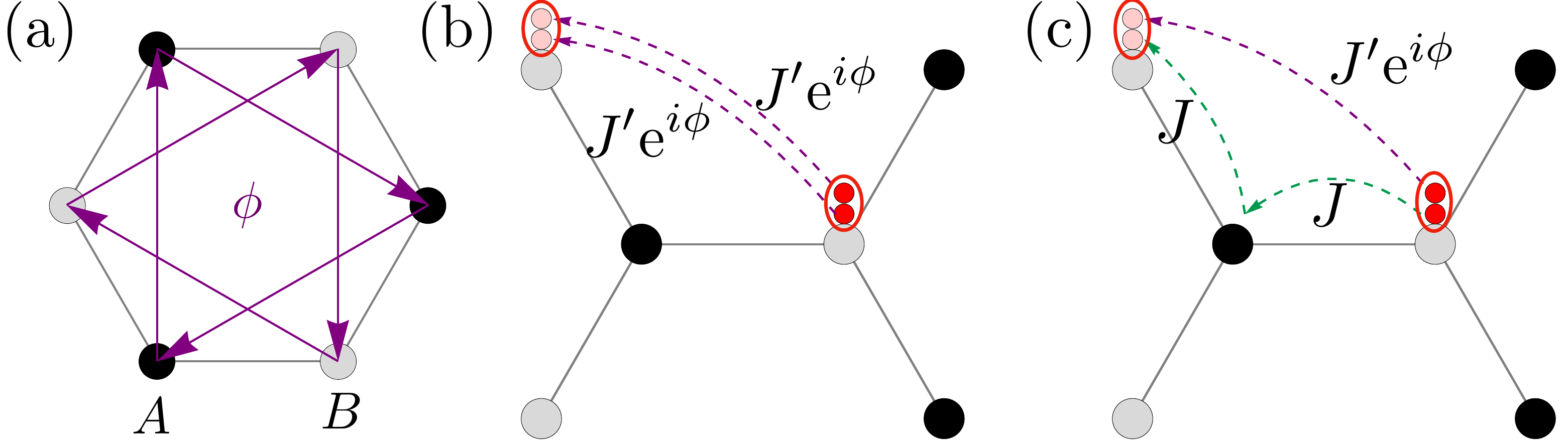}
\caption{ (a) Graphic representation of the Haldane model in the honeycomb lattice. The black and grey dots represents the $A$ and $B$ sites forming the two different sublattices. 
The phase $\phi$ of the next-nearest-neighbour hopping is taken to be positive along the directions set by the purple arrows. (b-c) An example of a next-nearest-neighbour hopping process for the doublon at second (b) and third (c) order in perturbation theory.}
\label{fig:system}
\end{figure}


\section{The model} 
\label{sec:model}
Starting from  the \textit{standard} Haldane model on the honeycomb lattice \cite{Haldane1988,Bernevig}
\begin{align}
&\mathcal{H}_H=- J \sum_{\langle\mathbf{r}^\prime,\mathbf{r}\rangle} \Big(\hat{a}_{\mathbf{r}^\prime}^\dagger \hat{b}^{}_{\mathbf{r}} + \text{H.c.}\Big) +
\frac{\Delta}{2} \sum_{\mathbf{r}\in A} \hat{a}_{\mathbf{r}}^\dagger \hat{a}^{}_{\mathbf{r}} 
-\frac{\Delta}{2} \sum_{\mathbf{r}\in B} \hat{b}_{\mathbf{r}}^\dagger \hat{b}^{}_{\mathbf{r}}
\notag
\\ & - J' \left( \sum_{{\llangle\mathbf{r}^\prime,\mathbf{r}\rrangle_{A}^+} }
\hat{a}_{\mathbf{r}'}^\dagger \hat{a}_{\mathbf{r}}^{}
\e^{i \phi}
+\sum_{\llangle\mathbf{r}^\prime,\mathbf{r}\rrangle_{B}^+}
\hat{b}_{\mathbf{r}'}^\dagger\hat{b}^{}_{\mathbf{r}}\e^{i \phi} + \text{H.c.}\right)\,,
\label{HNNN}
\end{align}
we introduce an on-site interaction term
\begin{equation}\mathcal{H}_U = \frac{U}{2}  \sum_{\mathbf{r}\in A}
 \hat{a}_{\mathbf{r}}^\dagger \hat{a}_{\mathbf{r}}^\dagger \hat{a}^{}_{\mathbf{r}} \hat{a}^{}_{\mathbf{r}}  +
\frac{U}{2}  \sum_{\mathbf{r} \in B}
   \hat{b}_{\mathbf{r}}^\dagger \hat{b}_{\mathbf{r}}^\dagger \hat{b}^{}_{\mathbf{r}} \hat{b}^{}_{\mathbf{r}}\,,
\label{Hinteraction}
\end{equation} 
and consider the two-body problem governed by the Hamiltonian $\mathcal{H}=\mathcal{H}_H+ \mathcal{H}_U$.  In our notations, $\hat{a}^{(\dagger)}_{\mathbf{r}}$ and $\hat{b}^{(\dagger)}_{\mathbf{r}}$ are the annihilation (creation) operators of a boson
at position $\mathbf{r}\in A,B $ either in the $A$ or $B$-sublattices.
The symbol ${\langle\mathbf{r'},\mathbf{r}\rangle}$ represents all nearest-neighbouring sites where $\mathbf{r}'$ and $\mathbf{r}$ belong to sublattice $A$ and $B$ respectively, whereas ${\llangle\mathbf{r'},\mathbf{r}\rrangle_{S}^{+}}$, with $S=A,B$, accounts for all next-nearest-neighbouring hopping processes taking place in the direction of the arrows in Fig.~\ref{fig:system}(a) within the same sublattice.   
The nearest-neighbour hopping coefficient $J>0$ is real, whereas the next-nearest-neighbour hopping coefficient is complex with modulus $J'$ and phase $\phi$. 
The on-site energy difference between the $A$ and $B$-sublattices is given by $\Delta$. In this work, we are going to focus on the repulsive $U>0$ case, which ensures stability in the ultracold atoms implementations. A similar analysis with analogous results can be carried out for $U<0$.


\section{Effective doublon Hamiltonian}  
\label{sec:effectiveHam}
For sufficiently large on-site interactions $U$, the high-energy sector of the spectrum describes a tightly-bound doublon state with energy of order $U$ \cite{Winkler2006, Strohmaier2010}, well separated from the scattering continuum. The decay into two free particles is energetically forbidden and the doublon is therefore stable. 
We can thus define the annihilation (creation) operators $\hat{\alpha}^{(\dagger)}_{\bf{r}}\equiv \hat a^{(\dagger)}_{\bf r} \hat a^{(\dagger)}_{\bf r}/\sqrt 2$ and $\hat{\beta}^{(\dagger)}_{\bf{r}}\equiv \hat b^{(\dagger)}_{\bf r} \hat b^{(\dagger)}_{\bf r}/\sqrt 2$ that destroy (create) a doublon 
at position $\mathbf{r}$ in the $A$ and $B$-sublattices, respectively. For large interactions, we define the doublon subspace spanned by $\{\hat{\alpha}^{\dagger}_{\bf{r}} |0\rangle, \hat{\beta}^{\dagger}_{\bf{r}} |0\rangle \}$.

The doublon dynamics can be described by an effective Hamiltonian in the doublon subspace and is obtained by treating hopping processes in perturbation theory, as depicted in Figs.~\ref{fig:system}(b-c).
Up to third order in perturbation theory, the doublon effective Hamiltonian takes the form
\begin{widetext}
\begin{align}
\label{Heff}
&\mathcal{H}_\text{eff} =  \\
& - J_\text{eff} \sum_{\langle \mathbf{r}^\prime,\mathbf{r}\rangle} \Big( \hat{\alpha}_{\mathbf{r}^\prime}^\dagger \hat{\beta}^{}_{\mathbf{r}} + \text{H.c.}\Big)  
-\left(\sum_{{\llangle \mathbf{r}^\prime,\mathbf{r}\rrangle}_A^+} J_\text{eff}'^A \e^{i \varphi_A}\, \hat{\alpha}_{\mathbf{r}^\prime}^\dagger \hat{\alpha}^{}_{\mathbf{r}}
-\sum_{{\llangle \mathbf{r}^\prime,\mathbf{r}\rrangle}_B^+} { J_\text{eff}'^{B}} \e^{i \varphi_B}\, \hat{\beta}_{\mathbf{r}^\prime}^\dagger \hat{\beta}^{}_{\mathbf{r}} + \text{H.c.}\right) 
+\sum_{\mathbf{r}\in A}  \Delta_{\text{eff}}^A \, \hat{\alpha}_{\mathbf{r}}^\dagger \hat{\alpha}^{}_{\mathbf{r}}  +\sum_{\mathbf{r}\in B} \Delta_{\text{eff}}^B \,\hat{\beta}_{\mathbf{r}}^\dagger \hat{\beta}^{}_{\mathbf{r}}\notag
\end{align}
\end{widetext}
where the full expression for the effective parameters is given in Appendix~\ref{sec:AppendixA}.

In the limit of very large interactions $U \ggg J,J',\Delta$, second order perturbation theory, which considers processes such as the one depicted in Fig.~\ref{fig:system}(b), is sufficient to well capture the main features of the system. 
The hopping parameters change sign compared to the single-particle case and take the simple form $J_{\textrm{eff}} = -2 J^2 U / (U^2-\Delta^2)$, $J'^{A,B}_{\textrm{eff}}= J'_{\infty}=-2J'^2/U$. In next-nearest neighbour hopping processes, doublons pick up a phase $\varphi_{A,B}=2\phi$ that is twice the single-particle one (see also Ref.~\cite{Bello2017}). The on-site energy difference instead becomes $\Delta^A_{\textrm{eff}}-\Delta^B_{\textrm{eff}} = 2\Delta + 12 J^2 U/(U^2-\Delta^2)$. Hence,  in the strongly interaction limit $U\ggg J$, second order processes recover a standard Haldane model with renormalized parameters.

The situation is even more interesting in the intermediate interaction regime ($U\approx 30 J$), which is relevant for experimental purposes \cite{Winkler2006,Greiner2017}, and where the presence of next-nearest-neighbour hopping processes makes the inclusion of third order corrections (as the one depicted in Fig.~\ref{fig:system}(c)) necessary for a quantitative description of the system. First of all, as one can see from the explicit formulas in Appendix~\ref{sec:AppendixA}, this leads to next-nearest-neighbour doublon hopping phases $\varphi_{A,B}\neq 2\phi$, namely different from twice the single-particle one.
Moreover, a non-zero on-site energy difference $\Delta$ between $A$ and $B$ sites makes the next-nearest-neighbour hopping coefficients of the effective Hamiltonian in Eq.~\eqref{Heff} different in the two sublattices both in amplitude $ J_\text{eff}'^A \neq J_\text{eff}'^B $ and phase $\varphi_A \neq \varphi_B$. 

As a consequence, doublons are effectively described by a \textit{generalized} Haldane model in Eq.~\eqref{Heff}, still belonging to the same topological class of the standard Haldane model. Therefore, we do expect non-trivial topological phases for the doublons as well. This is to be contrasted, for instance, with the SSH model in one dimension where on-site interactions break chiral symmetry, which is required for the topological protection of the edge states \cite{DiLiberto2016}.

\begin{figure}[t]
\centering
\includegraphics[width=\columnwidth]{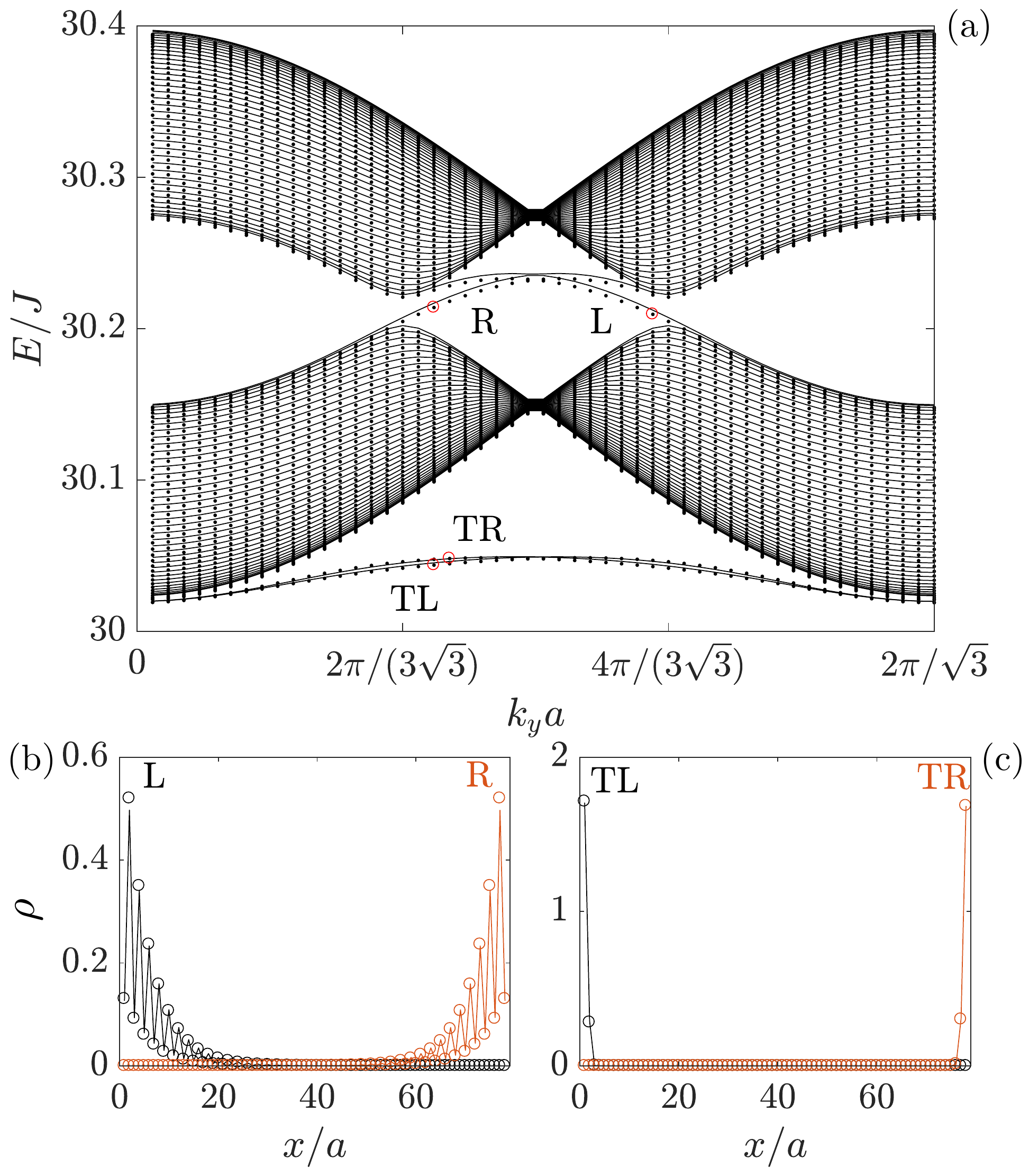}
\caption{(a) Doublon energy spectrum as a function of the center-of-mass momentum $k_y$, with periodic boundary conditions along $y$ and open boundary conditions along $x$ (bearded termination on both ends). The dots are obtained from an exact diagonalization of the two body Hamiltonian with the parameters $N_x=78$, $N_y=51$, $J'/J=0.2$, $U/J=30$, $\phi=\pi/5$, $\Delta=0$. The solid lines are obtained from the effective model in Eq.~\eqref{Heff} with the same parameters. 
(b-c) Particle density per site $\rho$ of the states highlighted with a red circle in (a), projected on the $x$-axis.
In (b) we show the topological states labelled with L and R, respectively, localized on the left (L) and on the right (R) edges.
In (c) we show the Tamm-Shockley states labelled with TL and TR, respectively, localized on the left (TL) and on the right (TR) edges.
Solid lines are obtained from the effective model whereas the open circles are the exact diagonalization results.}
\label{fig:Haldane1BB}
\end{figure}

\section{Doublon edge states} 
\label{sec:doublon}
To show the existence of topological states for doublons, we use the effective model in Eq.~\eqref{Heff} and exact diagonalization calculations in a ribbon geometry, with open boundary conditions in the $x$-direction and periodic boundary conditions in the $y$-direction. The center-of-mass momentum along $y$ is a good quantum number and takes the values $k_y=2\pi m/(\sqrt{3}a N_y)$, for $m=0,\dots N_y-1$, where $N_y$ is the number of sites in the $y$ direction and $a$ is the lattice spacing. 
A bearded type of termination is chosen on both edges along the $x$-direction, but analogous results are found also for zigzag terminations.

In Fig.~\ref{fig:Haldane1BB}(a), we show the doublon energy spectrum as a function of the center-of-mass momentum $k_y$. Dots are obtained from the exact diagonalization of the two-body Hamiltonian, while solid lines are calculated from the effective model in Eq.~\eqref{Heff}, with effective parameters calculated up to third order. The scattering continuum is at much lower energy and is not shown.

With open boundary conditions, the different connectivity of bulk and edge sites affects the contributions of virtual processes, leading to different bulk and edge parameters in the effective model in Eq.~\eqref{Heff} ~\cite{Bello2016,DiLiberto2016}. The main effect is to shift the on-site energy of the outermost sites, thus effectively changing the termination for the doublons from a bearded to a zigzag (see Appendix~\ref{sec:AppendixA} and~\ref{sec:AppendixB} for more details). 

The doublon spectrum shown in Fig.~\ref{fig:Haldane1BB}(a) is gapped, and presents two different types of edge states. In the gap between the two bands, we find topological edge states (R,L), corresponding to a zigzag type of termination \cite{Bernevig, CastroNeto2009}.
Moreover, two-body Tamm-Shockley edge states (TR,TL) \cite{Tamm1932,Shockley1939}, which are absent in the non-interacting limit, appear in the lower part of the spectrum.  The Tamm-Shockley states have no topological origin but are purely generated by interactions as a consequence of the different renormalization of the tight-binding parameters at the edges~\cite{DiLiberto2016,Bello2016,Gorlach2017}. Indeed, once the difference in the edge parameters is compensated by a suitable external potential, the Tamm-Shockley states disappear whereas the in-gap states are preserved, as shown in Appendix.~\ref{sec:AppendixB}.

In Fig.~\ref{fig:Haldane1BB}(b), we show the particle density per site of the topological edge states (L,R) projected along the $x$-direction. We clearly see that these states are exponentially localized on the left and right edges of the system, respectively. Figure~\ref{fig:Haldane1BB}(c) instead shows the particle density of the Tamm-Shockley edge states (TL,TR). The figures show a very good agreement between exact diagonalization (dots) and effective theory (solid line).

\begin{figure}[t]
\centering
\includegraphics[width=\columnwidth]{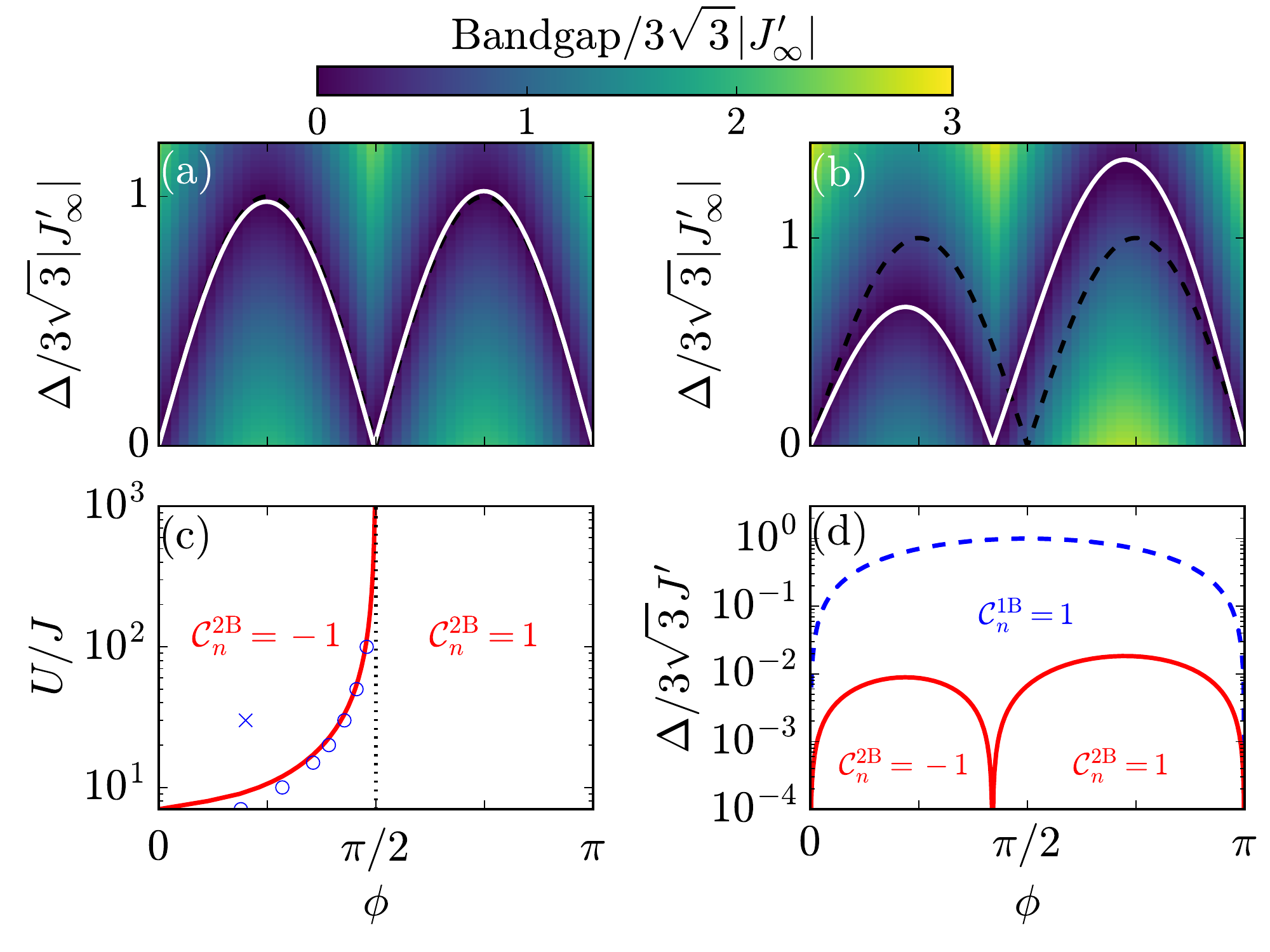}
\caption{(a)-(b) Gap between the two doublon bands as a function of the next-nearest-neighbour hopping phase $\phi$ and the onsite $A-B$ imbalance $\Delta$ in units of $ |J'_\infty | =2J'^2/U$ for $J'/J=0.2$, $U/J=500$ (a) and $U/J=30$ (b). The diagram is obtained from the exact diagonalization of a system with periodic boundary conditions ($N_x=60$ and $N_y=51$). The gap closing points track the topological phase transition. This transition is compared with the predictions of the phase boundaries at second order (black dashed line) and third order (white solid line) perturbation theory. 
(c) Critical phase at $\Delta=0$ as a function of $U$. Open circles are the gap closing points estimated with exact diagonalization. The dashed and the solid line correspond to the prediction of the effective model at second and third order, respectively. The cross corresponds to the parameters shown in Fig.~\ref{fig:Haldane1BB}. 
(d) Comparison between the single-particle topological phase diagram (blue dashed line) and the doublon topological phase diagram (red solid line) in units of the single-particle next-nearest-neighbour hopping $J'$, indicating the regions of non-vanishing Chern numbers.} 
\label{fig:bandgap_plot}
\end{figure}

\section{Topological phase diagram} 
\label{sec:topo}
To further characterize the topological properties of the doublons, we calculate the gap closing points in the parameter space for a system with periodic boundary conditions in both $x$ and $y$-direction, thus tracing the topological phase boundaries.

It is instructive to start from the very strongly-interacting limit $U \ggg J,J',\Delta$. In Fig.~\ref{fig:bandgap_plot}(a), we show the band gap for $U/J = 500$ calculated with exact diagonalization on a finite system. Moreover, we consider the effective model in Eq.~\eqref{Heff} up to second order, corresponding to a standard Haldane model with renormalized parameters, and in particular a next-nearest-neighbour hopping phase of $2\phi$. In this case, the phase boundaries are known analytically to be $\Delta_\text{eff}^{\textrm{crit}} = \pm 6\sqrt{3} \frac{J'^2}{U} \sin(2\phi)$, and are shown by the black dashed line in Fig.~\ref{fig:bandgap_plot}(a). From a comparison of this line with the prediction of the generalized Haldane model at third-order perturbation theory (white solid line) and the exact numerical results (color scale), we observe that higher-order processes are irrelevant at this value of $U$.
The Chern numbers for the different bands are calculated in momentum space from the effective model and are in agreement with the chirality of the edge states numerically obtained from exact diagonalization with open boundary conditions. As a consequence of the doubled next-nearest-neighbour hopping phase discussed above, we observe a two-lobe structure, with a staggered pattern of Chern number with period $\pi$. This is to be contrasted with the non-interacting Haldane model, where the phase boundaries and the Chern numbers display a $2\pi$ periodicity in $\phi$ \cite{Haldane1988}.
Moreover, as a consequence of the sign change of the effective doublon hopping $J_\text{eff}$ in Eq.~\eqref{Heff}, the staggered pattern of doublon Chern numbers for $\phi>0$ starts with $\mathcal C^{\textrm{2B}}_n =-1$, whereas it starts with $\mathcal C^{\textrm{1B}}_n =+1$ for the single-particle.

As the strength of interactions is reduced, third order processes become relevant. Even for $\Delta = 0$, the next-nearest-neighbour hopping phases $\varphi_{A,B}$ become complicated functions of the bare single-particle phase $\phi$, yielding the deformed lobes shown in Fig.~\ref{fig:bandgap_plot}(b).
Our results show that the doublon topological phase with edge states having the same chirality as the single-particle topological edge states dominates when interactions are reduced.
In Fig.~\ref{fig:bandgap_plot}, we only show a portion of the phase diagram. 
The phase boundaries are symmetric for $\Delta \rightarrow -\Delta$ and $\phi \rightarrow -\phi$, with the two lobes closer to $\phi=0$ becoming smaller for decreasing interactions. The alternating pattern of Chern numbers, that is the sequence $[-1,+1,-1,+1]$ in the interval $\phi \in [-\pi, \pi]$, is the same as the one discussed above in the strongly-interacting limit.
By comparison with exact-diagonalization, in Fig.~\ref{fig:bandgap_plot}(c) we observe that third order corrections quantitatively capture the boundaries of the topological transition down to $U/J \approx 15$.

\begin{figure*}[!ht]
\centering
\includegraphics[width=1.96 \columnwidth]{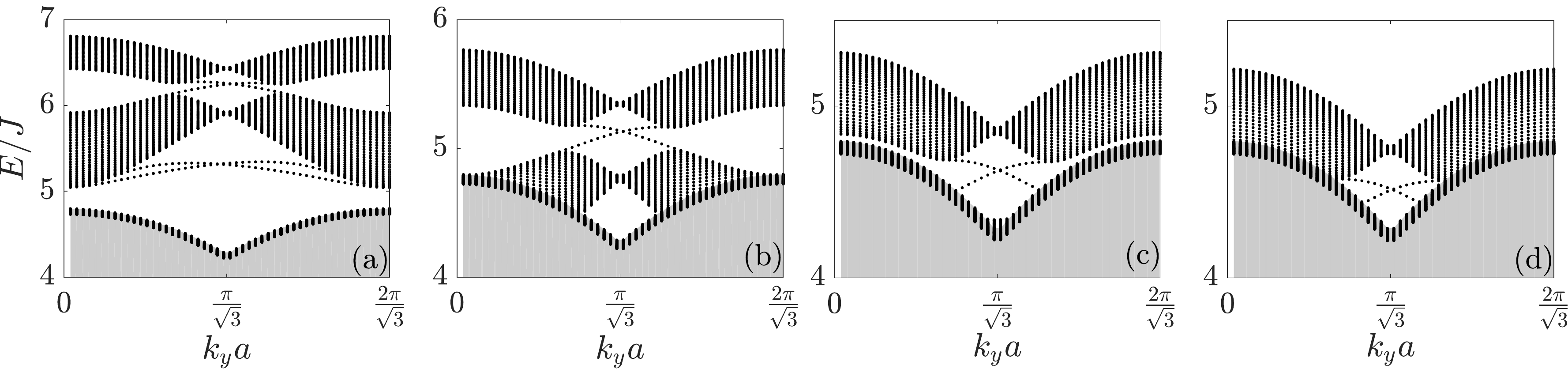}
\caption{(a)-(d) Energy spectra of the two-body model as a function of the center-of-mass momentum $k_y$ with different interactions, respectively $U/J=5,\,3.5,\,2.7,\,2.5$. The spectra are obtained from exact diagonalization of the two-body Hamiltonian with periodic boundary conditions along $y$ and open boundary conditions along $x$, with a bearded termination on both ends. Parameters are: $N_x=78$, $N_y=51$, $J'/J=0.2$, $\phi=\pi/3$, $\Delta/J=0.01$. Only the first $3 N_x$ eigenvalues from the top are shown. The grey area indicates the region of energy of the scattering continuum.}
\label{fig:ChangeU}
\end{figure*}

In Fig.~\ref{fig:bandgap_plot}(d), we summarize the doublon and single-particle phase diagram in units of the bare parameters. As expected, at large values of $\Delta$, both single-particles and doublons are in a non-topological phase. As the on-site energy difference $\Delta$ is reduced while keeping $\phi$ fixed, we first find a single-particle topological phase, whereas doublons remain topologically trivial. A topological phase transition for doublons only occurs when $\Delta$ is further reduced. This detrimental effect of interactions can be largely ascribed to the reduced next-nearest-neighbour hopping amplitude of the doublons.

The different regimes discussed above can be experimentally addressed by tuning the lattice parameters $\Delta$ and $\phi$ and the interaction strength $U$. This possibility opens several scenarios for experiments where topology is investigated through a dynamical protocol in either ultracold atoms \cite{Mancini2015,Greiner2017} or photon systems \cite{Google}. 
For given Hamiltonian parameters, one could prepare a pair of particles (atoms or photons) initially localized at the edge of a 2D system and compare the time-evolution of the density with the one of a chiral single-particle edge state. Doublons could manifest different types of behaviour: $(i)$ a propagation of a localized edge mode  with the same or opposite chirality as the single-particle state, $(ii)$ the spread into the bulk and the absence of a chiral signal.
This type of experiments would also highlight the presence of interaction-induced non-topological Tamm-Shockley edge states (see Fig. 2). Indeed, although an initial overlap with the Tamm-Shockley states affects the visibility of the chiral signal, the observation of an asymmetric propagation in the clockwise vs. counter-clockwise directions would give evidence of a superposition of topological and non-topological edge states.

\section{Small interactions}
\label{sec:smallU}

As a final point, we briefly discuss the regime of small interactions. The doublon bands are separated from the scattering bands of free particles approximately by the interaction energy $U$. As $U \approx J$, the doublon bands approach the scattering continuum. In this regime, it is natural to wonder about the fate of the two-body bound states and more specifically about the fate of the topological doublon edge states. 

This question is addressed in Fig.~\ref{fig:ChangeU}, where we show the upper part of the two-body spectrum, as a function of the center-of-mass momentum $k_y$ for different (decreasing) values of $U$. To avoid overcrowding the figure, we explicitly include only the highest scattering states, and indicate the rest of the scattering continuum with the shaded grey area.
We see that the doublon bands eventually touch the scattering continuum. Interestingly, we found that topological edge states are still present between the two upper bands down to $U=2.5 J$. Understanding this regime requires a further analysis to be carried out in future work.

\section{Conclusions}
\label{sec:conclusions}
We have investigated the physics of two interacting particles in the Haldane model as a function of the on-site interaction strength. We have found that, for given Hamiltonian parameters, the topological properties of the bound state can be very different from those of single-particle states. Most remarkably, in the intermediate interaction regime of experimental interest for state-of-the-art ultracold atom systems, the topological phase diagram of doublons is strongly influenced by their composite nature, as well their dynamics. Future work will address the consequences of the rich topological one- and two-body physics onto the collective properties of strongly correlated many-body systems of ultracold atoms~\cite{Bloch_rmp} or photons~\cite{Carusotto_rmp}.

\begin{acknowledgments}
We are grateful to T. Ozawa, H. M. Price, and M. Gorlach for useful discussions. This work was funded by the Autonomous Province of Trento, partially through the project SiQuro, by the EU--FET Proactive grant AQuS, Project No. 640800, and ERC Starting Grant TopoCold.
\end{acknowledgments}

\appendix
\section{Derivation of the effective Hamiltonian for the doublons}
\label{sec:AppendixA}

\begin{figure}[t]
\centering
\includegraphics[width=1 \columnwidth]{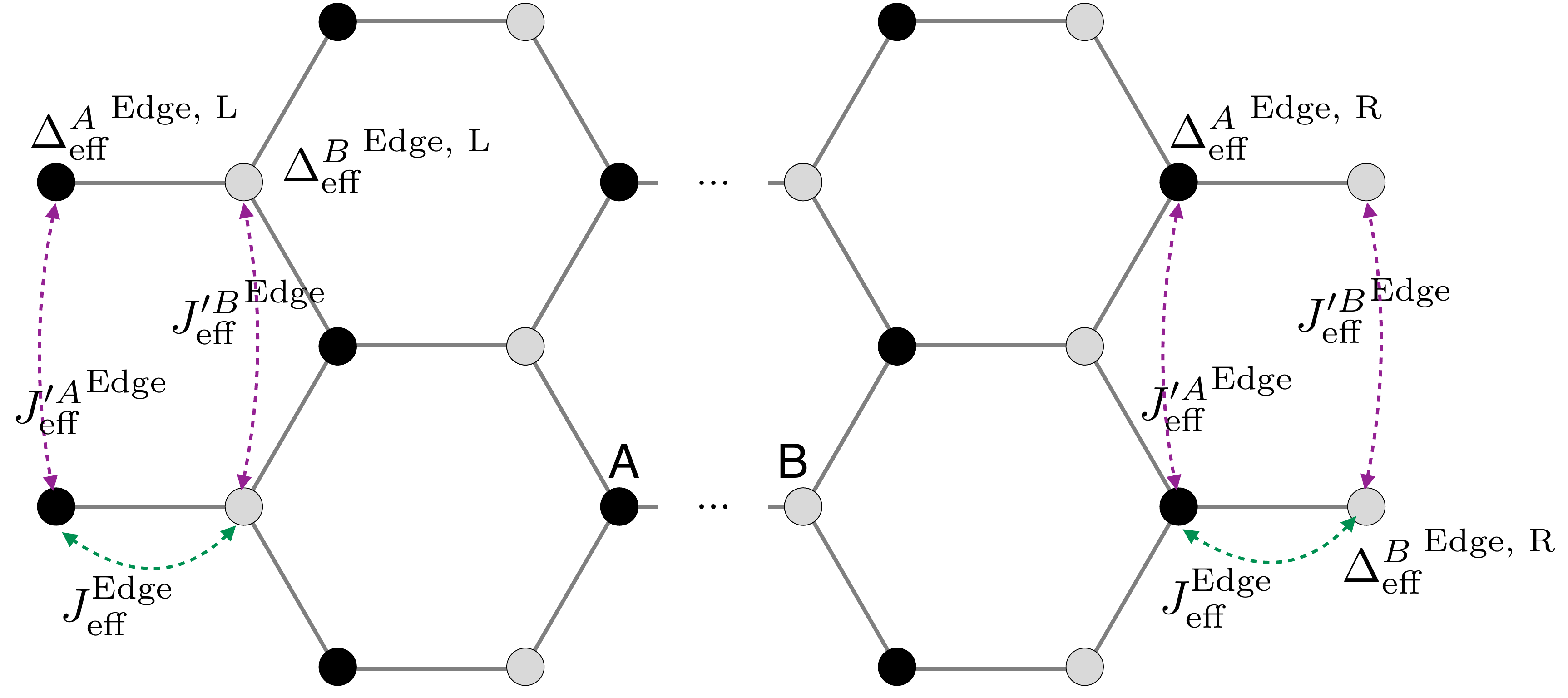}
\caption{A honeycomb lattice with bearded terminations on both ends. 
The on-site energy of the effective model for the doublon are specified for the left and the right edge, for both sublattices.}
\label{fig:EdgeBearded}
\end{figure}

In this first Appendix~\ref{sec:AppendixA}, we provide details on the derivation of the effective Hamiltonian for the two-body bound state in the strongly interacting regimes and we give explicit formulas for the next-to-leading order corrections that are needed to accurately describe the intermediate interaction regime via the generalized Haldane model of Eq.~\eqref{Heff}.  

We define the manifold of doublon states as the states $\ket{d}$ with double occupancy and obtain the effective Hamiltonian in the doublon subspace by accounting for the contributions of the single occupancy intermediate states $\ket{s}$ coupled to $\ket{d}$ by single-particle hopping processes. 
In the main text, we have defined the full Hamiltonian $\mathcal{H}=\mathcal{H}_H+ \mathcal{H}_U$ as the sum of the non-interacting Haldane model and the on-site interaction term in order to highlight the structure of the underlying single-particle model. For the purposes of the derivation of the effective Hamiltonian in the doublon subspace, it is instead useful to write $\mathcal{H}=\mathcal{H}_0+V$, being $\mathcal{H}_0$ the on-site part of the Hamiltonian and $V$ the nearest-neighbors and next-nearest-neighbors hopping terms, that we are going to treat as a perturbation in the strong interacting regime $U \gg J, \,J'$.

Hence, the on-site Hamiltonian $\mathcal{H}_0$, which can be treated exactly, is given by
\begin{align}
\label{unperturbed}
\mathcal{H}_0=&\frac{\Delta}{2} \sum_{\mathbf{r}\in A}  \hat{a}_{\mathbf{r}}^\dagger \hat{a}^{}_{\mathbf{r}}  - \frac{\Delta}{2} \sum_{\mathbf{r}\in B} \hat{b}_{\mathbf{r}}^\dagger \hat{b}^{}_{\mathbf{r}}\\
 &+\frac{U}{2}\sum_{\mathbf{r}\in A}\hat{a}_{\mathbf{r}}^\dagger \hat{a}_{\mathbf{r}}^\dagger \hat{a}^{}_{\mathbf{r}} \hat{a}^{}_{\mathbf{r}} + \frac{U}{2}\sum_{\mathbf{r}\in B} \hat{b}_{\mathbf{r}}^\dagger \hat{b}_{\mathbf{r}}^\dagger \hat{b}^{}_{\mathbf{r}} \hat{b}^{}_{\mathbf{r}}\,,\notag
\end{align}
while the perturbation $V$ reads
\begin{align}
\label{perturbed}
V=&- J \sum_{\langle \mathbf{r}^\prime,\mathbf{r}\rangle} \Big( \hat{a}_{\mathbf{r}^\prime}^\dagger \hat{b}^{}_{\mathbf{r}} + \text{H.c.} \Big)\\&- J' \left( \sum_{{\llangle \mathbf{r}^\prime,\mathbf{r}\rrangle}_A^+} \hat{a}_{\mathbf{r}^\prime}^\dagger \hat{a}^{}_{\mathbf{r}} \e^{i \phi}  +\sum_{{\llangle \mathbf{r}^\prime,\mathbf{r}\rrangle}_B^+}\hat{b}_{\mathbf{r}^\prime}^\dagger \hat{b}^{}_{\mathbf{r}}\e^{i \phi} +\text{H.c.}\right). \notag
\end{align}

As the perturbation $V$ only allows for single particle hopping processes, any matrix element that directly couples two double occupancy states is zero $\bra{d} V \ket{d'}=0$. It is therefore needed to include higher-order terms in perturbation theory. Going up to third order in $V$, the non-zero matrix elements of the effective Hamiltonian are given by \cite{Cohen, BirPikus}
\begin{widetext}
\begin{align}
\label{effective}
\bra{d} \mathcal{H}_\text{eff} \ket{d'} = &E_d^0 \delta_{dd'} +\frac{1}{2}\sum_s \bra{d} V \ket{s} \bra{s} V \ket{d'} \left[ \frac{1}{E_d^0-E_s^0}+\frac{1}{E_{d'}^0 - E_s^0 }\right]\\ 
&+\frac{1}{2}\sum_{s s'} \bra{d} V \ket{s} \bra{s} V \ket{s'} \bra{s'} V \ket{d'} \left[\frac{1}{\left(E_d^0-E_s^0\right)\left(E_d^0-E_{s'}^0\right)}+\frac{1}{\left(E_{d'}^0-E_s^0\right)\left(E_{d'}^0-E_{s'}^0\right)}\right],\notag
\end{align}
\end{widetext}
where $E_n^0$ are the eigenvalues of $\mathcal{H}_0$ relative to eigenstate $\ket{n}$. The matrix elements in Eq.~\eqref{effective} provide effective doublon hopping if 
$\ket{d} \neq \ket{d'}$ and effective energy shifts if $\ket{d} = \ket{d'}$.
Inserting explicitly Eq.~\eqref{unperturbed} and Eq.~\eqref{perturbed} in Eq.~\eqref{effective}, and reminding that doublon states $\ket{d}$ are spanned by the $\ket{\alpha_\mathbf{r}}$ and $\ket{\beta_\mathbf{r}}$ basis, we calculate the effective nearest-neighbour doublon hopping up to third order to be
\begin{align}
\label{JeffBulk}
-{J_\text{eff}} \equiv &\bra{\alpha_{\mathbf{r}^\prime}}\mathcal{H}_\text{eff} \ket{\beta_{\mathbf{r}}}= J^2 \left(\frac{1}{U+\Delta}+\frac{1}{U-\Delta}\right)\\
& -\frac{4 J^2 J' \cos(\phi)}{(U+\Delta)}\Bigg(\frac{1}{U} + \frac{1}{U+\Delta}+\frac{1}{U+2\Delta}\Bigg) \notag \\
&-\frac{4 J^2 J' \cos(\phi)}{(U-\Delta)}\Bigg(\frac{1}{U} + \frac{1}{U-\Delta}+\frac{1}{U-2\Delta}\Bigg).\notag 
\end{align}

In the same way, the effective next-nearest neighbour doublon hopping coefficients up to third order are 
\begin{align}
\label{JABulk}
- J_\text{eff}'^A &\e^{i \varphi_A} \equiv \bra{\alpha_{\mathbf{r}^\prime}}\mathcal{H}_\text{eff} \ket{\alpha_{\mathbf{r}}} =
\\& 2 \frac{J'^2 \e^{i 2\phi}}{U} -\frac{2J^2 J' \e^{i \phi}}{U+\Delta}\left(\frac{2}{U}+\frac{1}{U+\Delta}\right)-\frac{6 J'^3\e^{-i \phi}}{U^2}\,,\notag
\end{align}
and
\begin{align}
\label{JBBulk}
 -J_\text{eff}'^B & \e^{i \varphi_B} \equiv \bra{\beta_{\mathbf{r}^\prime}}\mathcal{H}_\text{eff} \ket{\beta_{\mathbf{r}}}=
\\&2 \frac{J'^2 \e^{i 2\phi}}{U} - \frac{2 J^2 J' \e^{i \phi}}{U-\Delta}\left(\frac{2}{U}+\frac{1}{U-\Delta}\right) -\frac{6 J'^3\e^{-i \phi}}{U^2}\,,\notag
\end{align}
where $\mathbf{r}^\prime$ and $\mathbf{r}$ lie within the same sublattice and hopping occurs in the positive direction defined by the arrows in Fig.~\ref{fig:system}(a) of the main text.

The doublon on-site energies on A and B sublattices are given by
\begin{align}
{\Delta_\text{eff}^A}&\equiv \bra{\alpha_{\mathbf{r}}}\mathcal{H}_\text{eff} \ket{\alpha_{\mathbf{r}}}=  U+ \Delta + \frac{6 J^2}{U+\Delta}
\\& + \frac{12 J'^2}{U} +\frac{24 J^2 J' \cos(\phi)}{U(U+\Delta)}+\frac{48 J'^3 \cos(3\phi)}{U^2} \notag 
\label{DeltaABulk}
\end{align}
and
\begin{align}
 {\Delta_\text{eff}^B} & \equiv \bra{\beta_{\mathbf{r}}}\mathcal{H}_\text{eff} \ket{\beta_{\mathbf{r}}} = U -\Delta + \frac{6 J^2}{U-\Delta}
 \\& + \frac{12 J'^2}{U} + \frac{24 J^2 J' \cos(\phi)}{U(U-\Delta)}+\frac{48 J'^3 \cos(3\phi)}{U^2}.\notag
\label{DeltaBBulk}
\end{align}
In Eqs.~\eqref{JeffBulk}-\eqref{DeltaBBulk}, the power of $J$ and $J'$ indicates how many nearest-neighbour and next-nearest-neighbour hopping processes are involved.

As long as $J'<J<U$ in Eqs.~\eqref{JeffBulk}-\eqref{DeltaBBulk}, there exists the hierarchy of energy scales $\frac{J'^3}{U^2}< \frac{J^2J'}{U^2}< \frac{J^2}{U}$, which makes only some of the contributions relevant for the parameters considered in the main text. In particular, the terms scaling like $J'^3$ will be generally negligible. However, a main ingredient of our analysis is provided by the competition between the first two terms in Eqs.~\eqref{JABulk}-\eqref{JBBulk}, which arises when interaction are reduced and the two ratios $\frac{J}{U}$ and $\frac{J'}{J}$ become comparable.

The hopping coefficients in Eqs.~\eqref{JeffBulk}-\eqref{JBBulk} and the on-site energies in Eqs.~\eqref{DeltaABulk}-\eqref{DeltaBBulk} are calculated by considering the coordination number of the lattice, and are therefore valid only in the bulk of the system. 
On a finite lattice, the lattice sites at the edges suffer some corrections with respect to the bulk in terms of effective hopping and on-site energy shifts, due to their different coordination number.
In Fig.~\ref{fig:EdgeBearded} we show all the quantities that get renormalized due to the presence of the edges.
In particular, the nearest-neighbour hopping on the last link of the bearded termination, for both left and  right edges, is modified as follows:
\begin{align}
-{J_\text{eff}}^{\text{Edge}} = &J^2 \left(\frac{1}{U+\Delta}+\frac{1}{U-\Delta}\right)\\
& -\frac{2 J^2 J' \cos(\phi)}{(U+\Delta)}\Bigg(\frac{1}{U} + \frac{1}{U+\Delta}+\frac{1}{U+2\Delta}\Bigg) \notag \\
&-\frac{2 J^2 J' \cos(\phi)}{(U-\Delta)}\Bigg(\frac{1}{U} + \frac{1}{U-\Delta}+\frac{1}{U-2\Delta}\Bigg).\notag
\end{align}
The next-nearest-neighbour hoppings on a vertical link are modified as:
\begin{align}
-{J_\text{eff}'^A}^{\text{Edge}}& e^{i \varphi_A}= -2 \frac{J'^2 \e^{i 2\phi}}{U} \\&
+\frac{2J^2 J' \e^{i \phi}}{U+\Delta}\left(\frac{2}{U}+\frac{1}{U+\Delta}\right)+\frac{3 J'^3\e^{-i \phi}}{U^2},\notag
\end{align}
and
\begin{align}
-{J_\text{eff}'^B}^{\text{Edge}} &e^{i \varphi_B}=-2 \frac{J'^2 \e^{i 2\phi}}{U} \\&
+ \frac{2 J^2 J' \e^{i \phi}}{U-\Delta}\left(\frac{2}{U}+\frac{1}{U-\Delta}\right) +\frac{3 J'^3\e^{-i\phi}}{U^2}.\notag
\end{align}
Instead, the values of the next-nearest-neighbour hoppings on diagonal links, which are not shown in Fig.~\ref{fig:EdgeBearded}, are the same as in the bulk.
The doublon on-site energies on a bearded termination on the left are:
\begin{align}
{\Delta_\text{eff}^A}^\text{Edge, L} &= U+ \Delta + \frac{2 J^2}{U+\Delta} + \frac{8 J'^2}{U} \\&
+ \frac{8 J^2 J' \cos(\phi)}{U(U+\Delta)}+\frac{24 J'^3 \cos(3\phi)}{U^2},\notag 
\end{align}
and
\begin{align}
{\Delta_\text{eff}^B}^\text{Edge, L} &= U- \Delta + \frac{6 J^2}{U-\Delta} + \frac{8 J'^2}{U} \\&
+ \frac{16 J^2 J' \cos(\phi)}{U(U-\Delta)}+\frac{24 J'^3 \cos(3\phi)}{U^2},\notag 
\end{align}
while on the right we have:
\begin{align}
{\Delta_\text{eff}^A}^\text{Edge, R} &= U+ \Delta + \frac{6 J^2}{U+\Delta} + \frac{8 J'^2}{U} \\&
+ \frac{16 J^2 J' \cos(\phi)}{U(U+\Delta)}+\frac{24 J'^3 \cos(3\phi)}{U^2},\notag 
\end{align}
and
\begin{align}
{\Delta_\text{eff}^B}^\text{Edge, R} &= U- \Delta + \frac{2 J^2}{U-\Delta} + \frac{8 J'^2}{U} \\&
+ \frac{8 J^2 J' \cos(\phi)}{U(U-\Delta)}+\frac{24 J'^3 \cos(3\phi)}{U^2}.\notag 
\end{align}

\begin{figure*}[t]
\centering
\includegraphics[width=1.95\columnwidth]{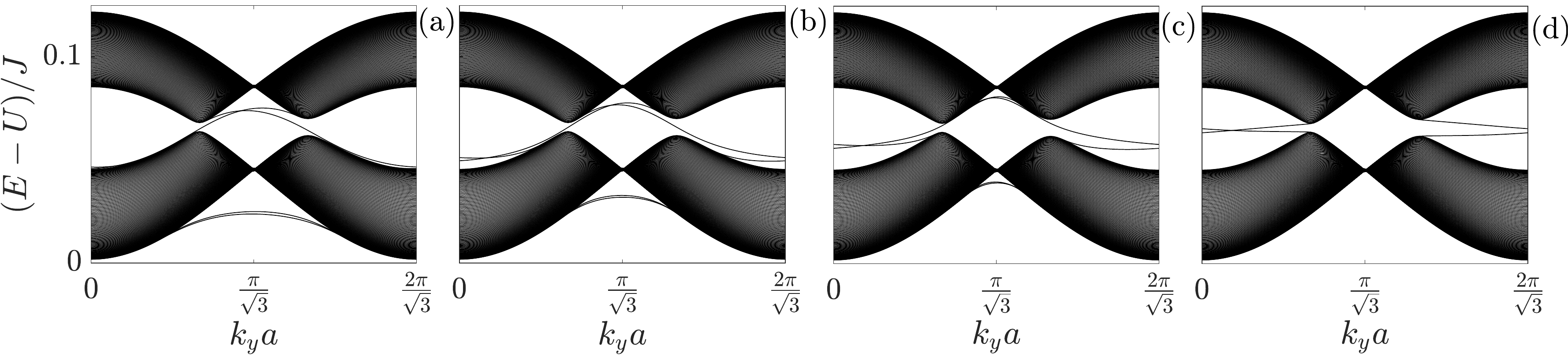}
\caption{Energy spectrum of the doublons as a function of the center-of-mass momentum $k_y$ along the $y$ direction, with periodic boundary conditions along $y$ and open boundary conditions along $x$ with a bearded termination on both ends. The spectra are obtained from the effective model in Eq.~\eqref{Heff} of the main text, for $N_x=100$, $J'/J=0.2$, $U/J=100$, $\phi=\pi/3$, $\Delta/J=0.001$ and with additional on-site energy shifts $\epsilon_\mu$ on the edge sites.
The energy of the site $\mu$ on the $\nu$ edge is increased by (a) $\epsilon_{\mu,\nu}= 0.25 \delta E_{\mu,\nu}$, (b) $\epsilon_{\mu,\nu}= 0.5 \delta E_{\mu,\nu}$, (c) $\epsilon_{\mu,\nu} = 0.75 \delta E_{\mu,\nu} $, and (d) $\epsilon_{\mu,\nu}= \delta E_{\mu,\nu} $, see notation in the text. The last case corresponds to the uniform situation where there is no energy difference between the edge sites and the bulk sites.}
\label{fig:Tamm}
\end{figure*} 

In particular, the on-site energies are different in the bulk and at the edges, with the edge sites having lower energy. We define the difference of on-site energy between bulk and edge as $\delta E_{\mu,\nu} = {\Delta_\text{eff}^{\mu}} - {\Delta_\text{eff}^\mu}^{\text{Edge, }\nu}$, where the index $\mu$ refers to A and B sublattices, while the index $\nu$ refers to the left or right edge.
The energy shift is the main cause of the non-topological localized Tamm-Shockley doublon states that appear at the edges \cite{Tamm1932,Shockley1939}, as discussed in the main text and in the following Appendix~\ref{sec:AppendixB}.

\section{Tamm-Shockley states}
\label{sec:AppendixB}

In this second Appendix~\ref{sec:AppendixB}, we provide more details on the Tamm-Shockley states that appear on the edge of the system and on their differences from topological edge states. 

This physics is illustrated in Fig.~\ref{fig:Tamm}, where we show the energy spectrum of the doublon as obtained from the effective model in Eq.~\eqref{Heff} of the main text. However, in addition to the $\delta E_{\mu,\nu}$ that arises from the effective model, we consider an arbitrary on-site potential $\epsilon_{\mu,\nu}$ acting on doublons sitting on the edge sites, where $\mu$ refers to A and B sublattices, while $\nu$ refers to the left or right edge. When the two energy shifts exactly compensate $\epsilon_{\mu,\nu}= \delta E_{\mu,\nu} $, we recover the uniform situation with no energy difference between edge and bulk sites.
The Tamm-Shockley states, which are labelled with (TL) and (TR) in Fig.~\ref{fig:Haldane1BB} of the main text, are present at the bottom of the lower band for $\epsilon_{\mu,\nu}=0$.
The energy difference $\delta E_{\mu,\nu}-\epsilon_{\mu,\nu}$ between bulk and edge sites is progressively decreased until it reaches zero, and the Tamm-Shockley states are continuously absorbed into the band. 

In the same figure, we also see that the presence of the additional edge versus bulk energy shift is progressively deforming the dispersion of the in-gap topological edge states, labelled with (L) and (R) in Fig.~\ref{fig:Haldane1BB} of the main text. 
In particular, we notice that the region of $k_y$ at which topological edge states are present gradually changes. For $\epsilon_{\mu,\nu}=0$, due to the onsite edge potential $\delta E_{\mu,\nu}$ on the last site of the bearded edge, the system effectively shows the topological edge properties of a zigzag termination. However, when the onsite edge potential $\delta E_{\mu,\nu}$ is fully compensated by the additional edge potential $\epsilon_{\mu,\nu}$, the edge properties are the usual ones corresponding to a bearded termination \cite{Hao2008, Bernevig,CastroNeto2009}.

It is important to remind that the existence of topological edge states follows from the non-triviality of the topological Chern invariant, which cannot change unless the bandgap closes. Since the bandgap cannot close as a result of an additional on-site edge potential, the edge states (L) and (R) are topologically protected. This is in contrast to what happens in the Su-Schrieffer-Heeger model, in which the renormalization of the effective parameters at the edges does not allow for the existence of topological doublon edge states \cite{DiLiberto2016, Gorlach2017, Bello2016, Bello2017}.


\begin{thebibliography}{56}
\expandafter\ifx\csname natexlab\endcsname\relax\def\natexlab#1{#1}\fi
\expandafter\ifx\csname bibnamefont\endcsname\relax
  \def\bibnamefont#1{#1}\fi
\expandafter\ifx\csname bibfnamefont\endcsname\relax
  \def\bibfnamefont#1{#1}\fi
\expandafter\ifx\csname citenamefont\endcsname\relax
  \def\citenamefont#1{#1}\fi
\expandafter\ifx\csname url\endcsname\relax
  \def\url#1{\texttt{#1}}\fi
\expandafter\ifx\csname urlprefix\endcsname\relax\def\urlprefix{URL }\fi
\providecommand{\bibinfo}[2]{#2}
\providecommand{\eprint}[2][]{\url{#2}}

\bibitem[{\citenamefont{Atala et~al.}(2013)\citenamefont{Atala, Aidelsburger,
  Barreiro, Abanin, Kitagawa, Demler, and Bloch}}]{Atala2013}
\bibinfo{author}{\bibfnamefont{M.}~\bibnamefont{Atala}},
  \bibinfo{author}{\bibfnamefont{M.}~\bibnamefont{Aidelsburger}},
  \bibinfo{author}{\bibfnamefont{J.~T.} \bibnamefont{Barreiro}},
  \bibinfo{author}{\bibfnamefont{D.}~\bibnamefont{Abanin}},
  \bibinfo{author}{\bibfnamefont{T.}~\bibnamefont{Kitagawa}},
  \bibinfo{author}{\bibfnamefont{E.}~\bibnamefont{Demler}}, \bibnamefont{and}
  \bibinfo{author}{\bibfnamefont{I.}~\bibnamefont{Bloch}},
  \bibinfo{journal}{Nat. Phys.} \textbf{\bibinfo{volume}{9}},
  \bibinfo{pages}{795} (\bibinfo{year}{2013}).

\bibitem[{\citenamefont{Aidelsburger et~al.}(2013)\citenamefont{Aidelsburger,
  Atala, Lohse, Barreiro, Paredes, and Bloch}}]{Aidelsburger2013}
\bibinfo{author}{\bibfnamefont{M.}~\bibnamefont{Aidelsburger}},
  \bibinfo{author}{\bibfnamefont{M.}~\bibnamefont{Atala}},
  \bibinfo{author}{\bibfnamefont{M.}~\bibnamefont{Lohse}},
  \bibinfo{author}{\bibfnamefont{J.~T.} \bibnamefont{Barreiro}},
  \bibinfo{author}{\bibfnamefont{B.}~\bibnamefont{Paredes}}, \bibnamefont{and}
  \bibinfo{author}{\bibfnamefont{I.}~\bibnamefont{Bloch}},
  \bibinfo{journal}{Phys. Rev. Lett.} \textbf{\bibinfo{volume}{111}},
  \bibinfo{pages}{185301} (\bibinfo{year}{2013}).

\bibitem[{\citenamefont{Miyake et~al.}(2013)\citenamefont{Miyake, Siviloglou,
  Kennedy, Burton, and Ketterle}}]{Miyake2013}
\bibinfo{author}{\bibfnamefont{H.}~\bibnamefont{Miyake}},
  \bibinfo{author}{\bibfnamefont{G.~A.} \bibnamefont{Siviloglou}},
  \bibinfo{author}{\bibfnamefont{C.~J.} \bibnamefont{Kennedy}},
  \bibinfo{author}{\bibfnamefont{W.~C.} \bibnamefont{Burton}},
  \bibnamefont{and} \bibinfo{author}{\bibfnamefont{W.}~\bibnamefont{Ketterle}},
  \bibinfo{journal}{Phys. Rev. Lett.} \textbf{\bibinfo{volume}{111}},
  \bibinfo{pages}{185302} (\bibinfo{year}{2013}).

\bibitem[{\citenamefont{Jotzu et~al.}(2014)\citenamefont{Jotzu, Messer,
  Desbuquois, Lebrat, Uehlinger, Greif, and Esslinger}}]{Jotzu2014}
\bibinfo{author}{\bibfnamefont{G.}~\bibnamefont{Jotzu}},
  \bibinfo{author}{\bibfnamefont{M.}~\bibnamefont{Messer}},
  \bibinfo{author}{\bibfnamefont{R.}~\bibnamefont{Desbuquois}},
  \bibinfo{author}{\bibfnamefont{M.}~\bibnamefont{Lebrat}},
  \bibinfo{author}{\bibfnamefont{T.}~\bibnamefont{Uehlinger}},
  \bibinfo{author}{\bibfnamefont{D.}~\bibnamefont{Greif}}, \bibnamefont{and}
  \bibinfo{author}{\bibfnamefont{T.}~\bibnamefont{Esslinger}},
  \bibinfo{journal}{Nature} \textbf{\bibinfo{volume}{515}},
  \bibinfo{pages}{237} (\bibinfo{year}{2014}).

\bibitem[{\citenamefont{Mancini et~al.}(2015)\citenamefont{Mancini, Pagano,
  Cappellini, Livi, Rider, Catani, Sias, Zoller, Inguscio, Dalmonte
  et~al.}}]{Mancini2015}
\bibinfo{author}{\bibfnamefont{M.}~\bibnamefont{Mancini}},
  \bibinfo{author}{\bibfnamefont{G.}~\bibnamefont{Pagano}},
  \bibinfo{author}{\bibfnamefont{G.}~\bibnamefont{Cappellini}},
  \bibinfo{author}{\bibfnamefont{L.}~\bibnamefont{Livi}},
  \bibinfo{author}{\bibfnamefont{M.}~\bibnamefont{Rider}},
  \bibinfo{author}{\bibfnamefont{J.}~\bibnamefont{Catani}},
  \bibinfo{author}{\bibfnamefont{C.}~\bibnamefont{Sias}},
  \bibinfo{author}{\bibfnamefont{P.}~\bibnamefont{Zoller}},
  \bibinfo{author}{\bibfnamefont{M.}~\bibnamefont{Inguscio}},
  \bibinfo{author}{\bibfnamefont{M.}~\bibnamefont{Dalmonte}},
  \bibnamefont{et~al.}, \bibinfo{journal}{Science}
  \textbf{\bibinfo{volume}{349}}, \bibinfo{pages}{1510} (\bibinfo{year}{2015}).

\bibitem[{\citenamefont{Stuhl et~al.}(2015)\citenamefont{Stuhl, Lu, Aycock,
  Genkina, and Spielman}}]{Stuhl2015}
\bibinfo{author}{\bibfnamefont{B.~K.} \bibnamefont{Stuhl}},
  \bibinfo{author}{\bibfnamefont{H.-I.} \bibnamefont{Lu}},
  \bibinfo{author}{\bibfnamefont{L.~M.} \bibnamefont{Aycock}},
  \bibinfo{author}{\bibfnamefont{D.}~\bibnamefont{Genkina}}, \bibnamefont{and}
  \bibinfo{author}{\bibfnamefont{I.~B.} \bibnamefont{Spielman}},
  \bibinfo{journal}{Science} \textbf{\bibinfo{volume}{349}},
  \bibinfo{pages}{1514} (\bibinfo{year}{2015}).

\bibitem[{\citenamefont{Hafezi et~al.}(2013)\citenamefont{Hafezi, Mittal, Fan,
  Migdall, and Taylor}}]{Hafezi2013}
\bibinfo{author}{\bibfnamefont{M.}~\bibnamefont{Hafezi}},
  \bibinfo{author}{\bibfnamefont{S.}~\bibnamefont{Mittal}},
  \bibinfo{author}{\bibfnamefont{J.}~\bibnamefont{Fan}},
  \bibinfo{author}{\bibfnamefont{A.}~\bibnamefont{Migdall}}, \bibnamefont{and}
  \bibinfo{author}{\bibfnamefont{J.~M.} \bibnamefont{Taylor}},
  \bibinfo{journal}{Nat. Photon.} \textbf{\bibinfo{volume}{7}},
  \bibinfo{pages}{1001} (\bibinfo{year}{2013}).

\bibitem[{\citenamefont{Rechtsman et~al.}(2013)\citenamefont{Rechtsman, Zeuner,
  Plotnik, Lumer, Podolsky, Dreisow, Nolte, Segev, and
  Szameit}}]{Rechtsman2013}
\bibinfo{author}{\bibfnamefont{M.~C.} \bibnamefont{Rechtsman}},
  \bibinfo{author}{\bibfnamefont{J.~M.} \bibnamefont{Zeuner}},
  \bibinfo{author}{\bibfnamefont{Y.}~\bibnamefont{Plotnik}},
  \bibinfo{author}{\bibfnamefont{Y.}~\bibnamefont{Lumer}},
  \bibinfo{author}{\bibfnamefont{D.}~\bibnamefont{Podolsky}},
  \bibinfo{author}{\bibfnamefont{F.}~\bibnamefont{Dreisow}},
  \bibinfo{author}{\bibfnamefont{S.}~\bibnamefont{Nolte}},
  \bibinfo{author}{\bibfnamefont{M.}~\bibnamefont{Segev}}, \bibnamefont{and}
  \bibinfo{author}{\bibfnamefont{A.}~\bibnamefont{Szameit}},
  \bibinfo{journal}{Nature} \textbf{\bibinfo{volume}{496}},
  \bibinfo{pages}{196} (\bibinfo{year}{2013}).

\bibitem[{\citenamefont{Chen et~al.}(2014)\citenamefont{Chen, Jiang, Chen, Zhu,
  Zhou, Dong, and Chan}}]{Chen2014}
\bibinfo{author}{\bibfnamefont{W.-J.} \bibnamefont{Chen}},
  \bibinfo{author}{\bibfnamefont{S.-J.} \bibnamefont{Jiang}},
  \bibinfo{author}{\bibfnamefont{X.-D.} \bibnamefont{Chen}},
  \bibinfo{author}{\bibfnamefont{B.}~\bibnamefont{Zhu}},
  \bibinfo{author}{\bibfnamefont{L.}~\bibnamefont{Zhou}},
  \bibinfo{author}{\bibfnamefont{J.-W.} \bibnamefont{Dong}}, \bibnamefont{and}
  \bibinfo{author}{\bibfnamefont{C.~T.} \bibnamefont{Chan}},
  \bibinfo{journal}{Nat. Comm.} \textbf{\bibinfo{volume}{5}},
  \bibinfo{pages}{5782} (\bibinfo{year}{2014}).

\bibitem[{\citenamefont{Ningyuan et~al.}(2015)\citenamefont{Ningyuan, Owens,
  Sommer, Schuster, and Simon}}]{Ningyuan2015}
\bibinfo{author}{\bibfnamefont{J.}~\bibnamefont{Ningyuan}},
  \bibinfo{author}{\bibfnamefont{C.}~\bibnamefont{Owens}},
  \bibinfo{author}{\bibfnamefont{A.}~\bibnamefont{Sommer}},
  \bibinfo{author}{\bibfnamefont{D.}~\bibnamefont{Schuster}}, \bibnamefont{and}
  \bibinfo{author}{\bibfnamefont{J.}~\bibnamefont{Simon}},
  \bibinfo{journal}{Phys. Rev. X} \textbf{\bibinfo{volume}{5}},
  \bibinfo{pages}{021031} (\bibinfo{year}{2015}).

\bibitem[{\citenamefont{Wang et~al.}(2009)\citenamefont{Wang, Chong,
  Joannopoulos, and Soljacic}}]{Wang2009}
\bibinfo{author}{\bibfnamefont{Z.}~\bibnamefont{Wang}},
  \bibinfo{author}{\bibfnamefont{Y.}~\bibnamefont{Chong}},
  \bibinfo{author}{\bibfnamefont{J.~D.} \bibnamefont{Joannopoulos}},
  \bibnamefont{and} \bibinfo{author}{\bibfnamefont{M.}~\bibnamefont{Soljacic}},
  \bibinfo{journal}{Nature} \textbf{\bibinfo{volume}{461}},
  \bibinfo{pages}{772} (\bibinfo{year}{2009}).

\bibitem[{\citenamefont{S{\"u}sstrunk and Huber}(2015)}]{Susstrunk2015}
\bibinfo{author}{\bibfnamefont{R.}~\bibnamefont{S{\"u}sstrunk}}
  \bibnamefont{and} \bibinfo{author}{\bibfnamefont{S.~D.} \bibnamefont{Huber}},
  \bibinfo{journal}{Science} \textbf{\bibinfo{volume}{349}},
  \bibinfo{pages}{47} (\bibinfo{year}{2015}).

\bibitem[{\citenamefont{Nash et~al.}(2015)\citenamefont{Nash, Kleckner, Read,
  Vitelli, Turner, and Irvine}}]{Nash2015}
\bibinfo{author}{\bibfnamefont{L.~M.} \bibnamefont{Nash}},
  \bibinfo{author}{\bibfnamefont{D.}~\bibnamefont{Kleckner}},
  \bibinfo{author}{\bibfnamefont{A.}~\bibnamefont{Read}},
  \bibinfo{author}{\bibfnamefont{V.}~\bibnamefont{Vitelli}},
  \bibinfo{author}{\bibfnamefont{A.~M.} \bibnamefont{Turner}},
  \bibnamefont{and} \bibinfo{author}{\bibfnamefont{W.~T.~M.}
  \bibnamefont{Irvine}}, \bibinfo{journal}{Proceedings of the National Academy
  of Sciences} \textbf{\bibinfo{volume}{112}}, \bibinfo{pages}{14495}
  (\bibinfo{year}{2015}).

\bibitem[{\citenamefont{Wang et~al.}(2015)\citenamefont{Wang, Lu, and
  Bertoldi}}]{Wang2015}
\bibinfo{author}{\bibfnamefont{P.}~\bibnamefont{Wang}},
  \bibinfo{author}{\bibfnamefont{L.}~\bibnamefont{Lu}}, \bibnamefont{and}
  \bibinfo{author}{\bibfnamefont{K.}~\bibnamefont{Bertoldi}},
  \bibinfo{journal}{Phys. Rev. Lett.} \textbf{\bibinfo{volume}{115}},
  \bibinfo{pages}{104302} (\bibinfo{year}{2015}).

\bibitem[{\citenamefont{He et~al.}(2016)\citenamefont{He, Ni, Ge, Sun, Chen,
  Lu, Liu, and Chen}}]{He2016}
\bibinfo{author}{\bibfnamefont{C.}~\bibnamefont{He}},
  \bibinfo{author}{\bibfnamefont{X.}~\bibnamefont{Ni}},
  \bibinfo{author}{\bibfnamefont{H.}~\bibnamefont{Ge}},
  \bibinfo{author}{\bibfnamefont{X.-C.} \bibnamefont{Sun}},
  \bibinfo{author}{\bibfnamefont{Y.-B.} \bibnamefont{Chen}},
  \bibinfo{author}{\bibfnamefont{M.-H.} \bibnamefont{Lu}},
  \bibinfo{author}{\bibfnamefont{X.-P.} \bibnamefont{Liu}}, \bibnamefont{and}
  \bibinfo{author}{\bibfnamefont{Y.-F.} \bibnamefont{Chen}},
  \bibinfo{journal}{Nat. Phys.} \textbf{\bibinfo{volume}{12}},
  \bibinfo{pages}{1124} (\bibinfo{year}{2016}).

\bibitem[{\citenamefont{Dalibard et~al.}(2011)\citenamefont{Dalibard, Gerbier,
  Juzeli\ifmmode~\bar{u}\else \={u}\fi{}nas, and \"Ohberg}}]{Dalibard2011}
\bibinfo{author}{\bibfnamefont{J.}~\bibnamefont{Dalibard}},
  \bibinfo{author}{\bibfnamefont{F.}~\bibnamefont{Gerbier}},
  \bibinfo{author}{\bibfnamefont{G.}~\bibnamefont{Juzeli\ifmmode~\bar{u}\else
  \={u}\fi{}nas}}, \bibnamefont{and}
  \bibinfo{author}{\bibfnamefont{P.}~\bibnamefont{\"Ohberg}},
  \bibinfo{journal}{Rev. Mod. Phys.} \textbf{\bibinfo{volume}{83}},
  \bibinfo{pages}{1523} (\bibinfo{year}{2011}).

\bibitem[{\citenamefont{Aidelsburger et~al.}(2011)\citenamefont{Aidelsburger,
  Atala, Nascimb\`ene, Trotzky, Chen, and Bloch}}]{Aidelsburger2011}
\bibinfo{author}{\bibfnamefont{M.}~\bibnamefont{Aidelsburger}},
  \bibinfo{author}{\bibfnamefont{M.}~\bibnamefont{Atala}},
  \bibinfo{author}{\bibfnamefont{S.}~\bibnamefont{Nascimb\`ene}},
  \bibinfo{author}{\bibfnamefont{S.}~\bibnamefont{Trotzky}},
  \bibinfo{author}{\bibfnamefont{Y.-A.} \bibnamefont{Chen}}, \bibnamefont{and}
  \bibinfo{author}{\bibfnamefont{I.}~\bibnamefont{Bloch}},
  \bibinfo{journal}{Phys. Rev. Lett.} \textbf{\bibinfo{volume}{107}},
  \bibinfo{pages}{255301} (\bibinfo{year}{2011}).

\bibitem[{\citenamefont{Goldman et~al.}(2014)\citenamefont{Goldman,
  Juzeli{\=u}nas, {\"O}hberg, and Spielman}}]{Goldman2014}
\bibinfo{author}{\bibfnamefont{N.}~\bibnamefont{Goldman}},
  \bibinfo{author}{\bibfnamefont{G.}~\bibnamefont{Juzeli{\=u}nas}},
  \bibinfo{author}{\bibfnamefont{P.}~\bibnamefont{{\"O}hberg}},
  \bibnamefont{and} \bibinfo{author}{\bibfnamefont{I.~B.}
  \bibnamefont{Spielman}}, \bibinfo{journal}{Reports on Progress in Physics}
  \textbf{\bibinfo{volume}{77}}, \bibinfo{pages}{126401}
  (\bibinfo{year}{2014}).

\bibitem[{\citenamefont{Aidelsburger et~al.}(2017)\citenamefont{Aidelsburger,
  Nascimb\`ene, and Goldman}}]{Aidelsburger2017}
\bibinfo{author}{\bibfnamefont{M.}~\bibnamefont{Aidelsburger}},
  \bibinfo{author}{\bibfnamefont{S.}~\bibnamefont{Nascimb\`ene}},
  \bibnamefont{and} \bibinfo{author}{\bibfnamefont{N.}~\bibnamefont{Goldman}},
  \bibinfo{journal}{arXiv:1710.00851}  (\bibinfo{year}{2017}).

\bibitem[{\citenamefont{Tsui et~al.}(1982)\citenamefont{Tsui, Stormer, and
  Gossard}}]{Tsui1982}
\bibinfo{author}{\bibfnamefont{D.~C.} \bibnamefont{Tsui}},
  \bibinfo{author}{\bibfnamefont{H.~L.} \bibnamefont{Stormer}},
  \bibnamefont{and} \bibinfo{author}{\bibfnamefont{A.~C.}
  \bibnamefont{Gossard}}, \bibinfo{journal}{Phys. Rev. Lett.}
  \textbf{\bibinfo{volume}{48}}, \bibinfo{pages}{1559} (\bibinfo{year}{1982}).

\bibitem[{\citenamefont{Laughlin}(1983)}]{Laughlin1983}
\bibinfo{author}{\bibfnamefont{R.~B.} \bibnamefont{Laughlin}},
  \bibinfo{journal}{Phys. Rev. Lett.} \textbf{\bibinfo{volume}{50}},
  \bibinfo{pages}{1395} (\bibinfo{year}{1983}).

\bibitem[{\citenamefont{S\o{}rensen et~al.}(2005)\citenamefont{S\o{}rensen,
  Demler, and Lukin}}]{Sorensen2005}
\bibinfo{author}{\bibfnamefont{A.~S.} \bibnamefont{S\o{}rensen}},
  \bibinfo{author}{\bibfnamefont{E.}~\bibnamefont{Demler}}, \bibnamefont{and}
  \bibinfo{author}{\bibfnamefont{M.~D.} \bibnamefont{Lukin}},
  \bibinfo{journal}{Phys. Rev. Lett.} \textbf{\bibinfo{volume}{94}},
  \bibinfo{pages}{086803} (\bibinfo{year}{2005}).

\bibitem[{\citenamefont{Hafezi et~al.}(2007)\citenamefont{Hafezi, S\o{}rensen,
  Demler, and Lukin}}]{Hafezi2007}
\bibinfo{author}{\bibfnamefont{M.}~\bibnamefont{Hafezi}},
  \bibinfo{author}{\bibfnamefont{A.~S.} \bibnamefont{S\o{}rensen}},
  \bibinfo{author}{\bibfnamefont{E.}~\bibnamefont{Demler}}, \bibnamefont{and}
  \bibinfo{author}{\bibfnamefont{M.~D.} \bibnamefont{Lukin}},
  \bibinfo{journal}{Phys. Rev. A} \textbf{\bibinfo{volume}{76}},
  \bibinfo{pages}{023613} (\bibinfo{year}{2007}).

\bibitem[{\citenamefont{Chen et~al.}(2012)\citenamefont{Chen, Gu, Liu, and
  Wen}}]{Chen2012}
\bibinfo{author}{\bibfnamefont{X.}~\bibnamefont{Chen}},
  \bibinfo{author}{\bibfnamefont{Z.-C.} \bibnamefont{Gu}},
  \bibinfo{author}{\bibfnamefont{Z.-X.} \bibnamefont{Liu}}, \bibnamefont{and}
  \bibinfo{author}{\bibfnamefont{X.-G.} \bibnamefont{Wen}},
  \bibinfo{journal}{Science} \textbf{\bibinfo{volume}{338}},
  \bibinfo{pages}{1604} (\bibinfo{year}{2012}).

\bibitem[{\citenamefont{Winkler et~al.}(2006)\citenamefont{Winkler, Thalhammer,
  Lang, Grimm, Hecker~Denschlag, Daley, Kantian, B{\"u}chler, and
  Zoller}}]{Winkler2006}
\bibinfo{author}{\bibfnamefont{K.}~\bibnamefont{Winkler}},
  \bibinfo{author}{\bibfnamefont{G.}~\bibnamefont{Thalhammer}},
  \bibinfo{author}{\bibfnamefont{F.}~\bibnamefont{Lang}},
  \bibinfo{author}{\bibfnamefont{R.}~\bibnamefont{Grimm}},
  \bibinfo{author}{\bibfnamefont{J.}~\bibnamefont{Hecker~Denschlag}},
  \bibinfo{author}{\bibfnamefont{A.~J.} \bibnamefont{Daley}},
  \bibinfo{author}{\bibfnamefont{A.}~\bibnamefont{Kantian}},
  \bibinfo{author}{\bibfnamefont{H.~P.} \bibnamefont{B{\"u}chler}},
  \bibnamefont{and} \bibinfo{author}{\bibfnamefont{P.}~\bibnamefont{Zoller}},
  \bibinfo{journal}{Nature} \textbf{\bibinfo{volume}{441}},
  \bibinfo{pages}{853} (\bibinfo{year}{2006}).

\bibitem[{\citenamefont{Valiente and Petrosyan}(2008)}]{Valiente2008}
\bibinfo{author}{\bibfnamefont{M.}~\bibnamefont{Valiente}} \bibnamefont{and}
  \bibinfo{author}{\bibfnamefont{D.}~\bibnamefont{Petrosyan}},
  \bibinfo{journal}{Journal of Physics B: Atomic, Molecular and Optical
  Physics} \textbf{\bibinfo{volume}{41}}, \bibinfo{pages}{161002}
  (\bibinfo{year}{2008}).

\bibitem[{\citenamefont{Valiente and Petrosyan}(2009)}]{Valiente2009}
\bibinfo{author}{\bibfnamefont{M.}~\bibnamefont{Valiente}} \bibnamefont{and}
  \bibinfo{author}{\bibfnamefont{D.}~\bibnamefont{Petrosyan}},
  \bibinfo{journal}{Journal of Physics B: Atomic, Molecular and Optical
  Physics} \textbf{\bibinfo{volume}{42}}, \bibinfo{pages}{121001}
  (\bibinfo{year}{2009}).

\bibitem[{\citenamefont{Mukherjee et~al.}(2016)\citenamefont{Mukherjee,
  Valiente, Goldman, Spracklen, Andersson, \"Ohberg, and
  Thomson}}]{Mukherjee2016}
\bibinfo{author}{\bibfnamefont{S.}~\bibnamefont{Mukherjee}},
  \bibinfo{author}{\bibfnamefont{M.}~\bibnamefont{Valiente}},
  \bibinfo{author}{\bibfnamefont{N.}~\bibnamefont{Goldman}},
  \bibinfo{author}{\bibfnamefont{A.}~\bibnamefont{Spracklen}},
  \bibinfo{author}{\bibfnamefont{E.}~\bibnamefont{Andersson}},
  \bibinfo{author}{\bibfnamefont{P.}~\bibnamefont{\"Ohberg}}, \bibnamefont{and}
  \bibinfo{author}{\bibfnamefont{R.~R.} \bibnamefont{Thomson}},
  \bibinfo{journal}{Phys. Rev. A} \textbf{\bibinfo{volume}{94}},
  \bibinfo{pages}{053853} (\bibinfo{year}{2016}).

\bibitem[{\citenamefont{Longhi and Valle}(2013)}]{Longhi2013}
\bibinfo{author}{\bibfnamefont{S.}~\bibnamefont{Longhi}} \bibnamefont{and}
  \bibinfo{author}{\bibfnamefont{G.~D.} \bibnamefont{Valle}},
  \bibinfo{journal}{Journal of Physics: Condensed Matter}
  \textbf{\bibinfo{volume}{25}}, \bibinfo{pages}{235601}
  (\bibinfo{year}{2013}).

\bibitem[{\citenamefont{Di~Liberto et~al.}(2016)\citenamefont{Di~Liberto,
  Recati, Carusotto, and Menotti}}]{DiLiberto2016}
\bibinfo{author}{\bibfnamefont{M.}~\bibnamefont{Di~Liberto}},
  \bibinfo{author}{\bibfnamefont{A.}~\bibnamefont{Recati}},
  \bibinfo{author}{\bibfnamefont{I.}~\bibnamefont{Carusotto}},
  \bibnamefont{and} \bibinfo{author}{\bibfnamefont{C.}~\bibnamefont{Menotti}},
  \bibinfo{journal}{Phys. Rev. A} \textbf{\bibinfo{volume}{94}},
  \bibinfo{pages}{062704} (\bibinfo{year}{2016}).

\bibitem[{\citenamefont{Di~Liberto et~al.}(2017)\citenamefont{Di~Liberto,
  Recati, Carusotto, and Menotti}}]{DiLiberto2017}
\bibinfo{author}{\bibfnamefont{M.}~\bibnamefont{Di~Liberto}},
  \bibinfo{author}{\bibfnamefont{A.}~\bibnamefont{Recati}},
  \bibinfo{author}{\bibfnamefont{I.}~\bibnamefont{Carusotto}},
  \bibnamefont{and} \bibinfo{author}{\bibfnamefont{C.}~\bibnamefont{Menotti}},
  \bibinfo{journal}{The European Physical Journal Special Topics}
  \textbf{\bibinfo{volume}{226}}, \bibinfo{pages}{2751} (\bibinfo{year}{2017}).

\bibitem[{\citenamefont{Gorlach and
  Poddubny}(2017{\natexlab{a}})}]{Gorlach2017}
\bibinfo{author}{\bibfnamefont{M.~A.} \bibnamefont{Gorlach}} \bibnamefont{and}
  \bibinfo{author}{\bibfnamefont{A.~N.} \bibnamefont{Poddubny}},
  \bibinfo{journal}{Phys. Rev. A} \textbf{\bibinfo{volume}{95}},
  \bibinfo{pages}{053866} (\bibinfo{year}{2017}{\natexlab{a}}).

\bibitem[{\citenamefont{Gorlach and
  Poddubny}(2017{\natexlab{b}})}]{Gorlach2017b}
\bibinfo{author}{\bibfnamefont{M.~A.} \bibnamefont{Gorlach}} \bibnamefont{and}
  \bibinfo{author}{\bibfnamefont{A.~N.} \bibnamefont{Poddubny}},
  \bibinfo{journal}{Phys. Rev. A} \textbf{\bibinfo{volume}{95}},
  \bibinfo{pages}{033831} (\bibinfo{year}{2017}{\natexlab{b}}).

\bibitem[{\citenamefont{Lim et~al.}(2011)\citenamefont{Lim, Troppenz, and
  Morais~Smith}}]{Lim2011}
\bibinfo{author}{\bibfnamefont{L.-K.} \bibnamefont{Lim}},
  \bibinfo{author}{\bibfnamefont{T.}~\bibnamefont{Troppenz}}, \bibnamefont{and}
  \bibinfo{author}{\bibfnamefont{C.}~\bibnamefont{Morais~Smith}},
  \bibinfo{journal}{Phys. Rev. A} \textbf{\bibinfo{volume}{84}},
  \bibinfo{pages}{053609} (\bibinfo{year}{2011}).

\bibitem[{\citenamefont{Bello et~al.}(2016)\citenamefont{Bello, Creffield, and
  Platero}}]{Bello2016}
\bibinfo{author}{\bibfnamefont{M.}~\bibnamefont{Bello}},
  \bibinfo{author}{\bibfnamefont{C.~E.} \bibnamefont{Creffield}},
  \bibnamefont{and} \bibinfo{author}{\bibfnamefont{G.}~\bibnamefont{Platero}},
  \bibinfo{journal}{Scientific Reports} \textbf{\bibinfo{volume}{6}},
  \bibinfo{pages}{22562} (\bibinfo{year}{2016}).

\bibitem[{\citenamefont{Bello et~al.}(2017)\citenamefont{Bello, Creffield, and
  Platero}}]{Bello2017}
\bibinfo{author}{\bibfnamefont{M.}~\bibnamefont{Bello}},
  \bibinfo{author}{\bibfnamefont{C.~E.} \bibnamefont{Creffield}},
  \bibnamefont{and} \bibinfo{author}{\bibfnamefont{G.}~\bibnamefont{Platero}},
  \bibinfo{journal}{Phys. Rev. B} \textbf{\bibinfo{volume}{95}},
  \bibinfo{pages}{094303} (\bibinfo{year}{2017}).

\bibitem[{\citenamefont{Marques and Dias}(2017{\natexlab{a}})}]{Marques2017}
\bibinfo{author}{\bibfnamefont{A.~M.} \bibnamefont{Marques}} \bibnamefont{and}
  \bibinfo{author}{\bibfnamefont{R.~G.} \bibnamefont{Dias}},
  \bibinfo{journal}{Phys. Rev. B} \textbf{\bibinfo{volume}{95}},
  \bibinfo{pages}{115443} (\bibinfo{year}{2017}{\natexlab{a}}).

\bibitem[{\citenamefont{Marques and Dias}(2017{\natexlab{b}})}]{Marques2017b}
\bibinfo{author}{\bibfnamefont{A.~M.} \bibnamefont{Marques}} \bibnamefont{and}
  \bibinfo{author}{\bibfnamefont{R.~G.} \bibnamefont{Dias}},
  \bibinfo{journal}{arXiv:1710.09148}  (\bibinfo{year}{2017}{\natexlab{b}}).

\bibitem[{\citenamefont{Qin et~al.}(2017)\citenamefont{Qin, Mei, Ke, Zhang, and
  Lee}}]{Lee1}
\bibinfo{author}{\bibfnamefont{X.}~\bibnamefont{Qin}},
  \bibinfo{author}{\bibfnamefont{F.}~\bibnamefont{Mei}},
  \bibinfo{author}{\bibfnamefont{Y.}~\bibnamefont{Ke}},
  \bibinfo{author}{\bibfnamefont{L.}~\bibnamefont{Zhang}}, \bibnamefont{and}
  \bibinfo{author}{\bibfnamefont{C.}~\bibnamefont{Lee}},
  \bibinfo{journal}{Phys. Rev. B} \textbf{\bibinfo{volume}{96}},
  \bibinfo{pages}{195134} (\bibinfo{year}{2017}).

\bibitem[{\citenamefont{Qin et~al.}(2016)\citenamefont{Qin, Mei, Ke, Zhang, and
  Lee}}]{Lee2}
\bibinfo{author}{\bibfnamefont{X.}~\bibnamefont{Qin}},
  \bibinfo{author}{\bibfnamefont{F.}~\bibnamefont{Mei}},
  \bibinfo{author}{\bibfnamefont{Y.}~\bibnamefont{Ke}},
  \bibinfo{author}{\bibfnamefont{L.}~\bibnamefont{Zhang}}, \bibnamefont{and}
  \bibinfo{author}{\bibfnamefont{C.}~\bibnamefont{Lee}},
  \bibinfo{journal}{arXiv:1611.00205}  (\bibinfo{year}{2016}).

\bibitem[{\citenamefont{Roushan et~al.}(2017)\citenamefont{Roushan, Neill,
  Megrant, Chen, Babbush, Barends, Campbell, Chen, Chiaro, Dunsworth
  et~al.}}]{Google}
\bibinfo{author}{\bibfnamefont{P.}~\bibnamefont{Roushan}},
  \bibinfo{author}{\bibfnamefont{C.}~\bibnamefont{Neill}},
  \bibinfo{author}{\bibfnamefont{A.}~\bibnamefont{Megrant}},
  \bibinfo{author}{\bibfnamefont{Y.}~\bibnamefont{Chen}},
  \bibinfo{author}{\bibfnamefont{R.}~\bibnamefont{Babbush}},
  \bibinfo{author}{\bibfnamefont{R.}~\bibnamefont{Barends}},
  \bibinfo{author}{\bibfnamefont{B.}~\bibnamefont{Campbell}},
  \bibinfo{author}{\bibfnamefont{Z.}~\bibnamefont{Chen}},
  \bibinfo{author}{\bibfnamefont{B.}~\bibnamefont{Chiaro}},
  \bibinfo{author}{\bibfnamefont{A.}~\bibnamefont{Dunsworth}},
  \bibnamefont{et~al.}, \bibinfo{journal}{Nat. Phys.}
  \textbf{\bibinfo{volume}{13}}, \bibinfo{pages}{146} (\bibinfo{year}{2017}).

\bibitem[{\citenamefont{Tai et~al.}(2017)\citenamefont{Tai, Lukin, Rispoli,
  Schittko, Menke, Borgnia, Preiss, Grusdt, Kaufman, and
  Greiner}}]{Greiner2017}
\bibinfo{author}{\bibfnamefont{M.~E.} \bibnamefont{Tai}},
  \bibinfo{author}{\bibfnamefont{A.}~\bibnamefont{Lukin}},
  \bibinfo{author}{\bibfnamefont{M.}~\bibnamefont{Rispoli}},
  \bibinfo{author}{\bibfnamefont{R.}~\bibnamefont{Schittko}},
  \bibinfo{author}{\bibfnamefont{T.}~\bibnamefont{Menke}},
  \bibinfo{author}{\bibfnamefont{D.}~\bibnamefont{Borgnia}},
  \bibinfo{author}{\bibfnamefont{P.~M.} \bibnamefont{Preiss}},
  \bibinfo{author}{\bibfnamefont{F.}~\bibnamefont{Grusdt}},
  \bibinfo{author}{\bibfnamefont{A.~M.} \bibnamefont{Kaufman}},
  \bibnamefont{and} \bibinfo{author}{\bibfnamefont{M.}~\bibnamefont{Greiner}},
  \bibinfo{journal}{Nature} \textbf{\bibinfo{volume}{546}},
  \bibinfo{pages}{519} (\bibinfo{year}{2017}).

\bibitem[{\citenamefont{Haldane}(1988)}]{Haldane1988}
\bibinfo{author}{\bibfnamefont{F.~D.~M.} \bibnamefont{Haldane}},
  \bibinfo{journal}{Phys. Rev. Lett.} \textbf{\bibinfo{volume}{61}},
  \bibinfo{pages}{2015} (\bibinfo{year}{1988}).

\bibitem[{\citenamefont{Bakr et~al.}(2009)\citenamefont{Bakr, Gillen, Peng,
  Folling, and Greiner}}]{Bakr2009}
\bibinfo{author}{\bibfnamefont{W.~S.} \bibnamefont{Bakr}},
  \bibinfo{author}{\bibfnamefont{J.~I.} \bibnamefont{Gillen}},
  \bibinfo{author}{\bibfnamefont{A.}~\bibnamefont{Peng}},
  \bibinfo{author}{\bibfnamefont{S.}~\bibnamefont{Folling}}, \bibnamefont{and}
  \bibinfo{author}{\bibfnamefont{M.}~\bibnamefont{Greiner}},
  \bibinfo{journal}{Nature} \textbf{\bibinfo{volume}{462}}, \bibinfo{pages}{74}
  (\bibinfo{year}{2009}).

\bibitem[{\citenamefont{Goldman et~al.}(2013)\citenamefont{Goldman, Dalibard,
  Dauphin, Gerbier, Lewenstein, Zoller, and Spielman}}]{Goldman2013}
\bibinfo{author}{\bibfnamefont{N.}~\bibnamefont{Goldman}},
  \bibinfo{author}{\bibfnamefont{J.}~\bibnamefont{Dalibard}},
  \bibinfo{author}{\bibfnamefont{A.}~\bibnamefont{Dauphin}},
  \bibinfo{author}{\bibfnamefont{F.}~\bibnamefont{Gerbier}},
  \bibinfo{author}{\bibfnamefont{M.}~\bibnamefont{Lewenstein}},
  \bibinfo{author}{\bibfnamefont{P.}~\bibnamefont{Zoller}}, \bibnamefont{and}
  \bibinfo{author}{\bibfnamefont{I.~B.} \bibnamefont{Spielman}},
  \bibinfo{journal}{Proceedings of the National Academy of Sciences}
  \textbf{\bibinfo{volume}{110}}, \bibinfo{pages}{6736} (\bibinfo{year}{2013}).

\bibitem[{\citenamefont{Goldman et~al.}(2016)\citenamefont{Goldman, Jotzu,
  Messer, G\"org, Desbuquois, and Esslinger}}]{Goldman2016}
\bibinfo{author}{\bibfnamefont{N.}~\bibnamefont{Goldman}},
  \bibinfo{author}{\bibfnamefont{G.}~\bibnamefont{Jotzu}},
  \bibinfo{author}{\bibfnamefont{M.}~\bibnamefont{Messer}},
  \bibinfo{author}{\bibfnamefont{F.}~\bibnamefont{G\"org}},
  \bibinfo{author}{\bibfnamefont{R.}~\bibnamefont{Desbuquois}},
  \bibnamefont{and}
  \bibinfo{author}{\bibfnamefont{T.}~\bibnamefont{Esslinger}},
  \bibinfo{journal}{Phys. Rev. A} \textbf{\bibinfo{volume}{94}},
  \bibinfo{pages}{043611} (\bibinfo{year}{2016}).

\bibitem[{\citenamefont{Bernevig and Hughes}(2013)}]{Bernevig}
\bibinfo{author}{\bibfnamefont{B.~A.} \bibnamefont{Bernevig}} \bibnamefont{and}
  \bibinfo{author}{\bibfnamefont{T.~L.} \bibnamefont{Hughes}},
  \emph{\bibinfo{title}{Topological insulators and topological
  superconductors}} (\bibinfo{publisher}{Princeton University Press},
  \bibinfo{year}{2013}).

\bibitem[{\citenamefont{Strohmaier et~al.}(2010)\citenamefont{Strohmaier,
  Greif, J\"ordens, Tarruell, Moritz, Esslinger, Sensarma, Pekker, Altman, and
  Demler}}]{Strohmaier2010}
\bibinfo{author}{\bibfnamefont{N.}~\bibnamefont{Strohmaier}},
  \bibinfo{author}{\bibfnamefont{D.}~\bibnamefont{Greif}},
  \bibinfo{author}{\bibfnamefont{R.}~\bibnamefont{J\"ordens}},
  \bibinfo{author}{\bibfnamefont{L.}~\bibnamefont{Tarruell}},
  \bibinfo{author}{\bibfnamefont{H.}~\bibnamefont{Moritz}},
  \bibinfo{author}{\bibfnamefont{T.}~\bibnamefont{Esslinger}},
  \bibinfo{author}{\bibfnamefont{R.}~\bibnamefont{Sensarma}},
  \bibinfo{author}{\bibfnamefont{D.}~\bibnamefont{Pekker}},
  \bibinfo{author}{\bibfnamefont{E.}~\bibnamefont{Altman}}, \bibnamefont{and}
  \bibinfo{author}{\bibfnamefont{E.}~\bibnamefont{Demler}},
  \bibinfo{journal}{Phys. Rev. Lett.} \textbf{\bibinfo{volume}{104}},
  \bibinfo{pages}{080401} (\bibinfo{year}{2010}).

\bibitem[{\citenamefont{Castro~Neto et~al.}(2009)\citenamefont{Castro~Neto,
  Guinea, Peres, Novoselov, and Geim}}]{CastroNeto2009}
\bibinfo{author}{\bibfnamefont{A.~H.} \bibnamefont{Castro~Neto}},
  \bibinfo{author}{\bibfnamefont{F.}~\bibnamefont{Guinea}},
  \bibinfo{author}{\bibfnamefont{N.~M.~R.} \bibnamefont{Peres}},
  \bibinfo{author}{\bibfnamefont{K.~S.} \bibnamefont{Novoselov}},
  \bibnamefont{and} \bibinfo{author}{\bibfnamefont{A.~K.} \bibnamefont{Geim}},
  \bibinfo{journal}{Rev. Mod. Phys.} \textbf{\bibinfo{volume}{81}},
  \bibinfo{pages}{109} (\bibinfo{year}{2009}).

\bibitem[{\citenamefont{Tamm}(1932)}]{Tamm1932}
\bibinfo{author}{\bibfnamefont{I.}~\bibnamefont{Tamm}}, \bibinfo{journal}{Phys.
  Z. Sowjetunion} \textbf{\bibinfo{volume}{1}}, \bibinfo{pages}{733}
  (\bibinfo{year}{1932}).

\bibitem[{\citenamefont{Shockley}(1939)}]{Shockley1939}
\bibinfo{author}{\bibfnamefont{W.}~\bibnamefont{Shockley}},
  \bibinfo{journal}{Phys. Rev.} \textbf{\bibinfo{volume}{56}},
  \bibinfo{pages}{317} (\bibinfo{year}{1939}).

\bibitem[{\citenamefont{Bloch et~al.}(2008)\citenamefont{Bloch, Dalibard, and
  Zwerger}}]{Bloch_rmp}
\bibinfo{author}{\bibfnamefont{I.}~\bibnamefont{Bloch}},
  \bibinfo{author}{\bibfnamefont{J.}~\bibnamefont{Dalibard}}, \bibnamefont{and}
  \bibinfo{author}{\bibfnamefont{W.}~\bibnamefont{Zwerger}},
  \bibinfo{journal}{Reviews of Modern Physics} \textbf{\bibinfo{volume}{80}},
  \bibinfo{pages}{885} (\bibinfo{year}{2008}).

\bibitem[{\citenamefont{Carusotto and Ciuti}(2013)}]{Carusotto_rmp}
\bibinfo{author}{\bibfnamefont{I.}~\bibnamefont{Carusotto}} \bibnamefont{and}
  \bibinfo{author}{\bibfnamefont{C.}~\bibnamefont{Ciuti}},
  \bibinfo{journal}{Reviews of Modern Physics} \textbf{\bibinfo{volume}{85}},
  \bibinfo{pages}{299} (\bibinfo{year}{2013}).

\bibitem[{\citenamefont{Cohen-Tannoudji
  et~al.}(1992)\citenamefont{Cohen-Tannoudji, Dupont-Roc, Grynberg, and
  Thickstun}}]{Cohen}
\bibinfo{author}{\bibfnamefont{C.}~\bibnamefont{Cohen-Tannoudji}},
  \bibinfo{author}{\bibfnamefont{J.}~\bibnamefont{Dupont-Roc}},
  \bibinfo{author}{\bibfnamefont{G.}~\bibnamefont{Grynberg}}, \bibnamefont{and}
  \bibinfo{author}{\bibfnamefont{P.}~\bibnamefont{Thickstun}},
  \emph{\bibinfo{title}{Atom-photon interactions: basic processes and
  applications}} (\bibinfo{publisher}{Wiley Online Library},
  \bibinfo{year}{1992}).

\bibitem[{\citenamefont{Bir and Pikus}(1974)}]{BirPikus}
\bibinfo{author}{\bibfnamefont{G.~L.} \bibnamefont{Bir}} \bibnamefont{and}
  \bibinfo{author}{\bibfnamefont{G.}~\bibnamefont{Pikus}},
  \emph{\bibinfo{title}{Summetry and strain-induced effects in semiconductors}}
  (\bibinfo{publisher}{Keter Publishing House Jerusalem},
  \bibinfo{year}{1974}).

\bibitem[{\citenamefont{Hao et~al.}(2008)\citenamefont{Hao, Zhang, Wang, Zhang,
  and Wang}}]{Hao2008}
\bibinfo{author}{\bibfnamefont{N.}~\bibnamefont{Hao}},
  \bibinfo{author}{\bibfnamefont{P.}~\bibnamefont{Zhang}},
  \bibinfo{author}{\bibfnamefont{Z.}~\bibnamefont{Wang}},
  \bibinfo{author}{\bibfnamefont{W.}~\bibnamefont{Zhang}}, \bibnamefont{and}
  \bibinfo{author}{\bibfnamefont{Y.}~\bibnamefont{Wang}},
  \bibinfo{journal}{Phys. Rev. B} \textbf{\bibinfo{volume}{78}},
  \bibinfo{pages}{075438} (\bibinfo{year}{2008}).

\end{thebibliography}

\end{document}